\documentclass[hidelinks,12pt]{article}
\usepackage[pdftex]{color, graphicx}
\usepackage{amsmath, amsfonts, amssymb, mathrsfs, mathtools}
\usepackage{dcolumn}
\usepackage{natbib}
\usepackage{caption}
\usepackage{listings}
\usepackage{subcaption}
\usepackage{algpseudocode}
\usepackage[]{algorithm}
\usepackage{algorithmicx}
\usepackage{stackengine}
\usepackage{multirow}
\usepackage{hyperref}

\usepackage{diagbox}
\usepackage{dsfont}
\usepackage{bm}

\bibpunct{(}{)}{;}{a}{}{,}
\newcommand{\x}{\mathbf{x}}
\newcommand{\X}{\mathbf{X}}
\newcommand{\Y}{\mathbf{Y}}
\newcommand{\K}{\mathbf{K}}
\newcommand{\kk}{\mathbf{k}}
\newcommand{\F}{\mathbb{F}}
\newcommand{\R}{\mathbb{R}}

\newcommand{\CC}{\mathcal{C}}
\newcommand{\GG}{\mathcal{G}}

\algrenewcommand{\algorithmiccomment}[1]{\hskip3em\#\# #1}
\algnewcommand{\LeftComment}[1]{\Statex \#\# #1}

\usepackage{setspace}

\begin{document}

	\title{\vspace{-1cm}
		Entropy-based adaptive design for \\contour finding and estimating reliability}
	\author{
		D.~Austin Cole\thanks{Corresponding author: \href{mailto:austin.cole8@vt.edu}{\tt austin.cole8@vt.edu}. GlaxoSmithKline, formerly at Virginia Tech}
		\and 
		Robert B.~Gramacy\thanks{Department of Statistics, Virginia Tech, Blacksburg, VA}
		\and James E.~Warner\thanks{NASA Langley, Hampton, VA}\footnotemark[3]
		\and Geoffrey F.~Bomarito\footnotemark[3]
		\and Patrick E.~Leser\footnotemark[3]
		\and William P.~Leser\footnotemark[3]
	}
	\date{}
	\maketitle

	\begin{abstract}
		In reliability analysis, methods used to estimate failure probability are
		often limited by the costs associated with model evaluations. Many of
		these methods, such as multifidelity importance sampling (MFIS), rely
		upon a computationally efficient, surrogate model like a Gaussian process
		(GP) to quickly generate predictions. The quality of the GP fit,
		particularly in the vicinity of the failure region(s), is instrumental in
		supplying accurately predicted failures for such strategies. We introduce
		an entropy-based GP adaptive design that, when paired with MFIS, provides
		more accurate failure probability estimates and with higher confidence. We
		show that our greedy data acquisition strategy better identifies multiple
		failure regions compared to existing contour-finding schemes. We then
		extend the method to batch selection, without sacrificing accuracy.
		Illustrative examples are provided on benchmark data as well as an
		application to an impact damage simulator for National Aeronautics and
		Space Administration (NASA) spacesuits.
	\end{abstract}
	
	
	\subsubsection*{Declarations:}
	The authors declare they have no conflict of interest. DAC recognizes support from the Engineering Research \& Analysis program at NASA.

	\newpage
	\section{Introduction}
	\label{sec:intro}
	
	Computer modeling of physical systems must accommodate uncertainty in
	materials and loading conditions. This input uncertainty translates into
	a stochastic response from the model, based on nominal settings of a
	physical system, even when the simulator is deterministic. In engineering,
	assessing the reliability of said system can mean guarding against a physical
	collapse, puncture or  failing of electronics.
	Reliability statistics like {\em failure probability}, the probability the
	response exceeds a threshold, can be calculated with Monte Carlo (MC) simulation. While
	MC produces an asymptotically unbiased estimator \citep{robert2013monte}, it
	can take thousands or even millions of model evaluations, i.e., great
	computational expense, to achieve a desired error tolerance.
	
	The search for alternatives to direct MC in computer-assisted reliability
	analysis has become a cottage industry of late. Some approaches seek to
	gradually reduce the design space for sampling through subset selection
	\citep{cannamela2008controlled, au2001estimation}. {\em Importance sampling}
	(IS) focuses MC efforts by biasing sampling toward areas of the design space
	where failure is probable \citep{srinivasan2013importance}, and then
	re-weights any expectations to correct for that bias asymptotically. Effective
	IS strategies  \citep{li2011efficient, peherstorfer2018multifidelity} aim to
	generate samples which reduce variance compared to direct MC.
	
	A class of {\em multifidelity} methods leverages a cheaper, low-fidelity
	computer or surrogate model \citep{fernandez2016review,
		peherstorfer2018survey, giselle2019issues}. Some variations seek to combine or
	hybridize information with the expensive high-fidelity analog
	\citep{giles2008multilevel, cliffe2011multilevel,litvinenko2013sampling}.
	Others adaptively trade off between sampling high- and low-fidelty models
	\citep{li2010evaluation, li2011efficient, bect2012sequential}. Multifidelity
	importance sampling \citep[MFIS;][]{peherstorfer2016multifidelity} taps a
	surrogate to derive a bias distribution for IS, preserving an estimator that
	is asymptotically unbiased while sparing computational resources. However, a
	surrogate that is inaccurate around the failure contour can wipe out any
	potential for computational gains.
	
	Gaussian process (GP) surrogates can help
	estimate failure probabilities through the experimental design technique known as {\em active learning} or {\em adaptive design}, described in Section \ref{ss:adaptive_design}. In
	the computer experiments community, many surrogate modelers are concerned with
	identifying {\em contours}, also known as level or excursion sets. Most
	strategies with GPs take a greedy design approach, with new data acquisitions
	optimizing a selection criteria.  Early adoptions modified the traditional
	Bayesian optimization criterion of expected improvement
	\citep{jones1998efficient} to sample around the contour
	\citep{bichon2008efficient, picheny2010adaptive, ranjan2008sequential}. Other
	criteria such as stepwise uncertainty reduction (SUR) aim to reduce the
	contour's uncertainty \citep{bect2012sequential,
		chevalier2014fast}. \cite{chevalier2013estimating} and
	\cite{azzimonti2016quantifying} use random set theory through the Vorob'ev
	expectation and deviation to minimize the posterior expected distance between
	the true and estimated contour.  Unfortunately, the best of these methods
	require numerical integration over the input space. Consequently most
	existing applications involve low-dimensional problems and/or focus on
	estimating contours that enclose a substantial volume of the design space.
	
	We are motivated by reliability applications in engineering that are often exemplified by expensive, high-fidelity simulators and potentially a large number of inputs (large input volume/space). These problems frequently include small failure probability, targeting a contour delineating a failure region
	occupying less than 0.1\% of that space. In addition, we desire a level of conservatism in our analysis. Though many problems contain a single failure region, being conservative includes guarding against missing one or more failure regions. 
	We found that a GP surrogate adaptively designed for contour finding, combined
	with MFIS, can provide more accurate reliability statistics, with less
	variability, for fixed sampling effort. Consequently, we introduce a new
	adaptive design acquisition function based on {\em entropy} that we call
	entropy-based contour locator (ECL). Although entropy has been leveraged for
	active learning with GP surrogates before
	\citep[e.g.][]{marques2018contour,oakley2004estimating}, our setup is
	distinct. We target more challenging response surfaces with more extreme
	contours, as motivated by our reliability context. Compared to alternatives,
	we find that our setup better explores the design space nearby the contours
	that outline {\em all} failure regions. Our criterion may be calculated in
	closed form because it does not rely upon numerical integration, allowing it
	to more easily scale to larger input dimensions. Acknowledging that many
	expensive simulation campaigns involve parallel evaluation on supercomputers,
	we extend our criterion to batch selection.
	
	\begin{figure}
		\includegraphics[trim=10 10 10 10, clip, width=\textwidth]{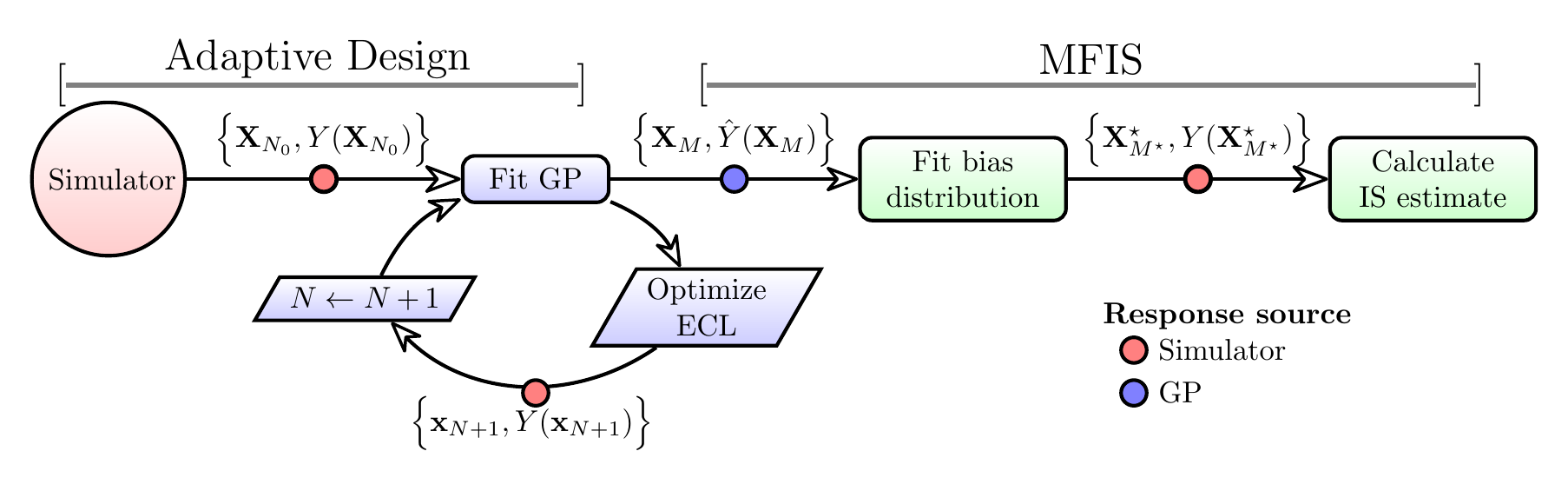}
		\caption{Full process combining ECL adaptive design with MFIS. Colored
			circles represent data with the response generated from the simulator (red)
			or GP (blue).
			\label{fig:flowchart}}
	\end{figure}
	
	The remainder of the paper proceeds as follows. In Section
	\ref{sec:background}, we provide the necessary background information
	regarding failure probability estimation, MFIS, and GP adaptive design. We
	introduce the ECL active learning strategy in Section \ref{sec:entropy}.
	Section \ref{sec:toy_results} showcases ECL empirically on a set of benchmark
	problems. Then we combine ECL designs with MFIS on a more involved and
	realistic, motivating application in Section \ref{sec:xemu_results}, involving
	a reliability analysis for a NASA spacesuit design under impact loading.
	Section \ref{sec:discussion} provides a concluding discussion.  Figure
	\ref{fig:flowchart} provides a flowchart for our setup.  Although some of the
	notation in that chart is not yet defined, we include it here because the flow
	mirrors the paper development as described above.  We encourage the reader to
	refer back to this figure as related concepts are introduced in Section
	\ref{sec:background}.  Our main contribution (Section \ref{sec:entropy})
	comprises the cycle in blue; however it is motivated by the green sections,
	which are showcased on NASA's spacesuit.

	\section{Reliability estimation tools}
	\label{sec:background}
	
	Here we provide the necessary context for estimating failure probability and contours.
	
	\subsection{Failure probability estimation}
	\label{ss:failure_prob}
	
	Consider an input--output system, modeled at high-fidelity with $\mathcal{T}:
	\mathcal{X} \equiv \R^d
	\rightarrow \R$ such that evaluations $\mathcal{T}(\x) \rightarrow y$ are
	computationally/time intensive. 
	Let $\mathbb{F}$ denote a distribution over $\x \in \mathcal{X}$, with density
	$f$, representing  uncertainties in the inputs $\x$. Reliability is defined
	through a \emph{limit state function}, $g:\mathbb{R} \rightarrow \mathbb{R}$,
	where failure is deemed to occur when $g(y) > 0$
	\citep{melchers2018structural}. In engineering, a failure is often
	characterized as an output $y$ exceeding a certain threshold $T$, expressing the limit state function as $g(y) = y
	- T$. The failure region is characterized as
	\begin{equation} 
		\label{eq:failure_region}
		\GG = \{\x\in \mathcal{X} :g(\mathcal{T}(\x))> 0\}, \quad \mbox{implying contour} \quad
		\CC = \{\x\in \mathcal{X} :g(\mathcal{T}(\x))= 0\}.
	\end{equation}
	With indicator function $\mathds{1}$, the probability of failure ($\alpha$) is defined as 
	\begin{equation} \label{eq:limit_state}
		\alpha=\mathbb{E}_{X}[g(\mathcal{T}(\x))>0]=\int_{\x\in \mathcal{X}}\mathds{1}_{\{g(\mathcal{T}(\x))>0\}} \; d \F. 
	\end{equation}
	For technical/theoretical aspects of development in sequential design/active
	learning literature, the contour of interest $\CC$ is sometimes assumed to be
	the {\em zero contour}, without loss of generality. However, our
	interest lies in identifying low/high quantiles of $\mathcal{T}(\x)$,
	corresponding to a small $\alpha$.  Our experience is that this setting is more challeging than zero or other more central contours in practice.  We 
	believe it is prudent to caution the reader against taking such liberties and arbitrarily describing $\CC$. 
	
	Direct MC sampling is perhaps the most common numerical integration scheme for
	Eq.~(\ref{eq:limit_state}), yielding approximate estimator
	$\hat{\alpha}_{\text{MC}}$. With $M$ samples, $\x_i
	\stackrel{\mathrm{iid}}{\sim}  \F$, $i=1, \dots, M$, the MC estimate for
	$\alpha$ and its associated variance are
	\begin{align} \label{eq:monte_carlo}
		\hat{\alpha}_{\text{MC}}&= \frac{1}{M}\sum_{i=1}^M \mathds{1}_{\{g(\mathcal{T}(\x_i))>0\}} & \mbox{with} &&
		\text{Var}\left(\hat{\alpha}_{\text{MC}}\right)&=\frac{\hat{\alpha}_{\text{MC}}\left(1-\hat{\alpha}_{\text{MC}}\right)}{M}.
	\end{align}
	While $\hat{\alpha}_{\text{MC}}$ in \eqref{eq:monte_carlo} is asymptotically
	unbiased \citep{robert2013monte}, it can take $M$ in the millions to achieve
	a desirable accuracy, which might tax resources required for evaluation of
	$\mathcal{T}$. For example, if $\alpha=10^{-4}$ and we wish to have a root mean
	squared error (RMSE) of $10^{-6}$ (corresponding to
	$\text{Var}\left(\hat{\alpha}_{\text{MC}}\right) < 10^{-12}$), $M > 10^8$ samples are needed.
	
	\label{ss:mfis}
	One solution to reducing this cost is through a focused simulation like {\em
		importance sampling} (IS). With IS, $\x^\star_i \stackrel{\mathrm{iid}}{\sim}
	\F_\star$ are generated from an auxiliary distribution biased toward
	region(s) of inputs more likely to cause failure (i.e., land in $\GG$). The
	resulting estimate
	\begin{equation} \label{eq:is_estimate}
		\hat{\alpha}_{\text{IS}}=P^{\text{IS}}(\x_1^\star,\dots,\x^\star_{M^\star})=\frac{1}{M^\star}\sum_{i=1}
		^{M^\star}\mathds{1}_{\{g(\mathcal{T}(\x_i^\star))>0\}}w(\x_i^\star) \quad \mbox{via weights} \quad
		w(\x^\star)=\frac{f(\x_i^\star)}{f_\star(\x_i^\star)},
	\end{equation}
	is asymptotically unbiased given $\text{supp}(f)\subseteq\text{supp}(f_\star)$
	\citep[][Section 14.2]{robert2013monte}. Constructing {\em bias distribution} $\F_\star$ can be
	expensive in its own right, motivating the search for a cheaper
	low-fidelity model, like a surrogate $\mathcal{S}$.  
	
	\emph{Multifidelity importance sampling} (MFIS) combines the low cost of
	generating samples from $\mathcal{S}$ when training  the bias
	distribution in IS \citep{peherstorfer2016multifidelity}. Algorithm
	\ref{alg:importance_sampling} shows pseudocode for MFIS: a bias
	distribution is generated by first evaluating $M$ samples with the cheap
	surrogate model $\mathcal{S}_N$, foreshadowing our notation for a GP surrogate
	(Section \ref{ss:adaptive_design}) which is trained on $N$ samples $\X_N \in
	\mathcal{X}$ paired with high-fidelity outputs $\Y_N$. Using the limit state
	function $g$, the samples that are classified to produce a failure response
	are used to train the bias distribution $\F_\star$ for IS.
	\cite{peherstorfer2016multifidelity} suggest using a Gaussian mixture model
	\citep[GMM;][]{celeux1995gaussian} for $\F_\star$. A set of $M^\star$ samples
	is drawn from $\F_\star$ and used to evaluate the high-fidelity model. The
	associated responses $\Y_{M^\star}$ are used along with IS weights
	$\mathbf{w}_{M^\star}$ to yield failure probability estimate
	$\hat{\alpha}^{\text{MFIS}}$.
	\begin{algorithm} \caption{Multifidelity Importance sampling} 
		\begin{algorithmic}[1] 
			\Procedure
			{MFIS}{$\mathcal{T}$, $\mathcal{S}_N$, $\F$, $\F_\star$, $M$, $M^\star$, $\kappa$}
			\State Draw $M$ samples $\X_{M}=\{\x_1,\dots,\x_{M}\}$ from $\F$
			\State $\hat{\Y}_{M} \leftarrow \mathcal{S}_N(\X_{M})$ \hfill \Comment{Predictions from surrogate}
			\State Train a GMM of $\kappa$ clusters with set $\mathcal{F}=\{\x\in\mathcal{X}:g(\hat{y}(\x))>0\}$\hfill \Comment{Form $\F_\star$}
			\State Draw $M^\star$ samples $\X^\star_{M^\star}=\{\x^\star_1, \dots,\x^\star_{M^\star}\}$ from $\F_\star$
			\State $\mathbf{w}_{M^\star}=\Big[\frac{f(\x^\star_1)}{f_\star(\x^\star_1)},\dots,\frac{f(\x^\star_{M^\star})}{f_\star(\x^\star_{M^\star})}\Big]$ \hfill \Comment{Compute importance weights}
			\State $\Y_{M^\star} \leftarrow \mathcal{T}(\X^\star_{M^\star})$ \hfill \Comment{Evaluate high-fidelity model}
			\State $\hat{\alpha}_\text{MFIS} =\frac{1}{M^\star}\sum_{i=1}
			^{M^\star}\mathbb{I}_{g(y_i)>0}w(\x^\star_i) $\hfill \Comment{MFIS estimate through \eqref{eq:is_estimate}}
			\State \Return $\hat{\alpha}_\text{MFIS}$
			\EndProcedure 
		\end{algorithmic} \label{alg:importance_sampling} 
	\end{algorithm}
	
	MFIS seeks to preserve the unbiasedness of the estimator
	\eqref{eq:is_estimate} while minimizing the number of high-fidelity
	evaluations $M^\star$. Yet the question of how to train a $\mathcal{S}_N$ for
	MFIS is not widely discussed in the literature. As a simple experiment, we conduct MFIS with the GP $\mathcal{S}_N$ on the Ishigami function
	\citep{ishigami1990importance}:
	$\mathcal{T}^{\text{Ish}}(\x)=\sin(x_1)+5\sin^2(x_2)+0.1x_3^4\sin(x_1)$ for
	$\x\in [-\pi, \pi]^3$. $\mathcal{T}^{\text{Ish}}$ is smooth, with six global
	minima creating six disjoint failure regions. We define the failure region
	$\GG^{\text{Ish}}$ for values $y$ below the threshold $T=-9.2244$. To align
	this with our definition in \eqref{eq:failure_region}, let $g^{\text{Ish}}(y)=
	- y+9.2244$, which corresponds to $\alpha^{\text{Ish}}\approx0.001$. In our
	experiment, we studied estimates of $\alpha^{\text{Ish}}$ for increasing
	budgets of simulator evaluations in an MC exercise with fifty repeats.
	
\begin{figure}
		\includegraphics[trim=20 0 0 0,clip,width=\textwidth]{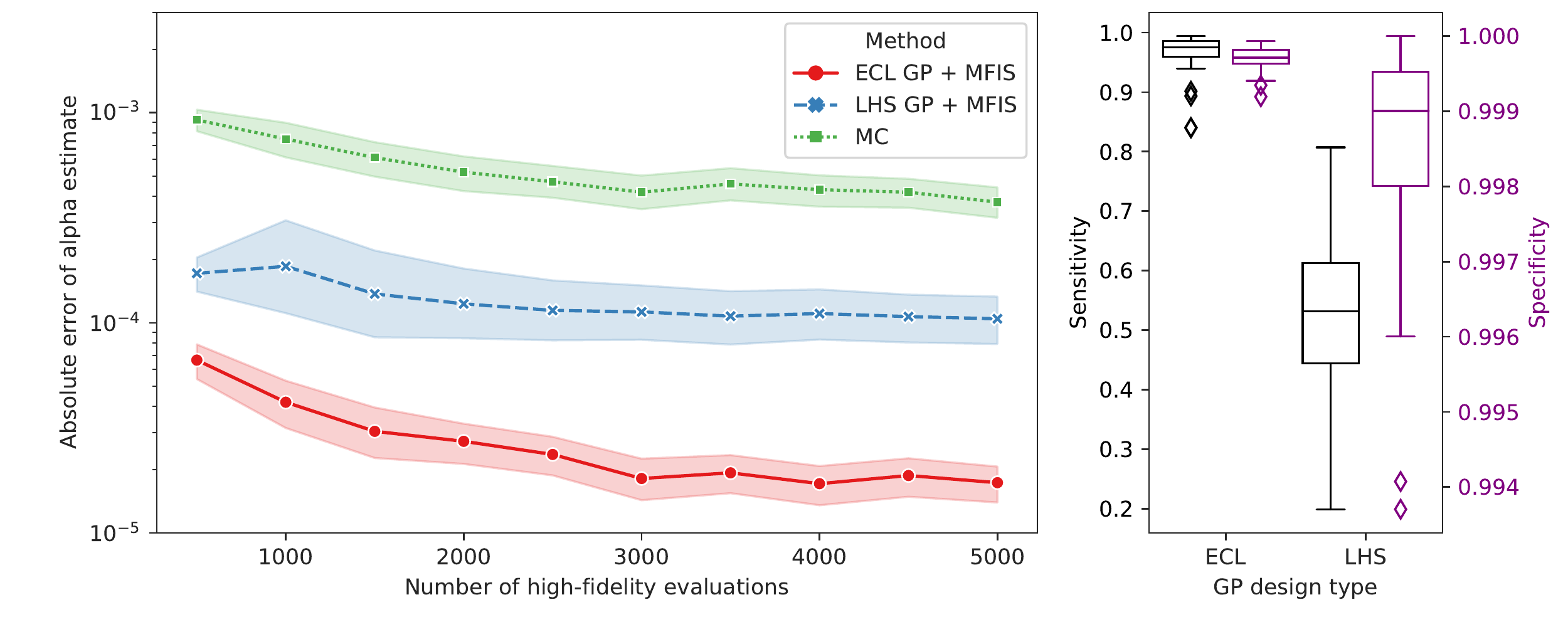}	
		\caption{{\em Left:} Absolute errors of failure estimates (truth is
			$10^{-4}$) for the Ishigami function. Bold lines denote mean error over 50
			samples with shading for the 95\% confidence regions. {\em Right:} failure
			region classification statistics across 50 MC repeats for the two GPs.
			\label{fig:ad_vs_sf_bakeoff}}
	\end{figure}

	The left panel of Figure \ref{fig:ad_vs_sf_bakeoff} shows errors
	$|\hat{\alpha}-\alpha^{\text{Ish}}|$ for three comparators with
	the dark lines indicating means, and 95\% confidence interval (CI) as shaded
	bands. A simple MC estimate $\hat{\alpha}_{\text{MC}}$ from \eqref{eq:monte_carlo}
	is along the green dotted line; observe that its absolute error decreases as the number of
	high-fidelity samples ($M$ on the $x$-axis) increases, showing a slow convergence to the truth (zero error). The blue dashed line represents MFIS-based estimates from \eqref{eq:is_estimate} using a GP $\mathcal{S}_N$ trained on $N=200$ space-filling Latin
	Hypercube samples \citep[LHS;][]{Mckay:1979} generating evaluations of the
	simulator $\mathcal{T}^{\text{Ish}}$, followed by $M^\star$ draws from
	$\F_\star$ (with $N+M^\star$ on the $x$-axis).  Finally, the red solid line represents the GP trained via our proposed entropy-based adaptive
	design (30 LHS followed by 170 acquisitions; details in Section
	\ref{sec:entropy}). Notice that the red solid line starts better than both (small
	$M^\star$) and, unlike the blue dashed line, offers noticeable further improvement (larger
	$M^\star$).
	
	The right panel of Figure \ref{fig:ad_vs_sf_bakeoff} shows the proportion of
	the input space that is correctly classified as being inside (black/left
	$y$-axis) and outside (magenta/right) the failure region.  This calculation is
	based on a  $10^6$-sized LHS in the 3d input space. Observe that the
	LHS-based GP produces considerably inferior estimates.  The adaptive
	method correctly identifies most of failure points (sensitivity) and
	nearly all of the non-failure points (specificity). GPs trained on
	space-filling LHSs often miss some of the failure regions, resulting in a much
	lower sensitivity. Lower specificity in the LHS option means that
	more locations that are not failures are being labeled as such. If used for
	MFIS, the result would be a flawed biasing, leading to inefficient
	reliability estimates.

	\subsection{Adaptively designed Gaussian processes surrogates}
	\label{ss:adaptive_design}
	
	GPs make popular surrogates due to their tremendous
	flexibility in modeling complex surfaces
	\citep{Sacks1989,Santer2018,gramacy2020surrogates}. They may be simply
	characterized by placing a multivariate normal (MVN) prior on the
	observations/outputs $\Y_N$. MVNs are uniquely defined by a mean vector $\mu$,
	often assumed to be zero in the GP regression context for simplicity, and an
	$N\times N$ covariance matrix $\K_N$. For $N$ observations included in
	$D_N=(\X_N, \Y_N)$, the joint model for the responses is $\Y_N\sim
	\mathcal{N}_N(\mathbf{0},\K_N)$. Entries of $\K_N$ are calculated from a kernel
	$k_{\Psi}(\x_i, \x_j)$, which is usually based on inverse
	distances in the input space. Many expressions of $k_{\Psi}(\cdot, \cdot)$ exist, including the squared exponential
	and Mat\'ern families \citep{Stein2012}. The kernel hyperparameters $\Psi$, such as $\tau^2$ (scale) and
	$\theta$ (lengthscale), are estimated by optimizing the MVN log-likelihood.

	The true power of a GP lies in its ability to produce accurate predictions by
	conditioning on observed data.  The predictive equations for a new testing
	location $\x'$ are:
	\begin{align} \label{eq:pred_eqs}
		\begin{split}
			\mu_N(\x'|D_N)&=\kk(\x', \X_N)\kk(\X_N, \X_N)^{-1}\Y_N \\
			\sigma_N^2(\x'|D_N)&= k_{\Psi}(\x',\x')-\kk(\x', \X_N)\kk(\X_N, \X_N)^{-1}\kk(\x', \X_N)^\top,
		\end{split}
	\end{align}
	where $\kk(\x',\X_N)$ is the vector of cross evaluations of the kernel $k_{\Psi}(\cdot, \cdot)$  between $\x'$ and $\X_N$.
	When we write $\mathcal{S}_N$ we are referring to these equations without a
	particular $\x'$ in mind. A GP's ability to produce fast, accurate
	predictions has fueled the rise of \emph{active learning} (i.e. sequential
	design) techniques in the machine learning community
	\citep{settles2011theories}. The idea is to leverage the current model's
	information (fit to existing data) to greedily select the next sample/batch of
	inputs used to gather more data, updating that information. Making such
	selections relies on the optimization of an \emph{acquisition function}. A
	litany of acquisition functions serve diverse modeling goals. These include
	minimizing predictive variance \citep{mackay1992information, Seo2000,
		cohn1994adavances}, finding function optima
	\citep{jones1998efficient,gramacy4027optimization, wu2017bayesian} via
	so-called Bayesian optimization (BO), and accurate hyperparameter optimization
	\citep{Gramacy2015, zhang2018distance}.
	
	\subsubsection*{Targeting level sets}
	
	Several acquisition functions target level sets, or {\em contours}, and are
	tailored to GP regression. For our purposes, the contour of interest is the
	boundary of the failure region(s) in \eqref{eq:failure_region}. Some acquisition functions focus
	squarely on sampling where $\CC$ is uncertain \citep{bect2012sequential,
		echard2010kriging, ranjan2008sequential, bichon2008efficient,
		lee2008sampling}. Others are inspired by BO solutions, striking a balance between
	extrema-seeking and global posterior uncertainty \citep{bryan2008actively,
		gotovos2013active, bogunovic2016advances}. When applying contour finding for
	failure probability estimation, it is vital to identify all disjoint
	regions within $\CC$ (when more than one exist). Therefore, it is highly desirable for an adaptive design
	method to promote exploration. Consequently many acquisition functions
	involve integrating over the domain $\mathcal{X}$ \citep{ranjan2008sequential,
		bichon2008efficient, picheny2010adaptive, chevalier2013estimating,
		chevalier2014fast}. We know of no examples where such integrals that can be evaluated analytically.
	Numerical schemes, based on reference grids and other 
	quadrature, do not scale well to large dimension $d$.  Some
	shortcuts are helpful, e.g., \cite{lyu2021evaluating} leverage predictive
	variance in \eqref{eq:pred_eqs} to simplify some aspects. However, challenges
	remain to produce a method tractable for estimating a small-volume contour
	in a big $\mathcal{X}$.
	
	A complimentary class of approaches involve identifying $\CC$ by mapping to a
	classification setting \citep{gotovos2013active, azzimonti2016quantifying,
		bolin2015excursion}. A common measure of uncertainty for discrete random
	variables is {\em entropy}. Originating from information theory
	\citep{cover2006elements}, the entropy of a discrete random variable $W$ is
	measured as
	\begin{equation} \label{eq:gen_entropy}
		H(W)=-\sum_{i=1}^k \mathbb{P}(w_i)\log \mathbb{P}(w_i),
	\end{equation}
	where $i={1,\dots,k}$ indexes events $w_i$, with probability mass
	$\mathbb{P}(w_i)$. A mapping that would apply in our setting, as a criterion
	for valuing inputs $\x$, is with $k=2$ where
	\begin{equation} \label{eq:ecl_categories}
		w_1(\x) \equiv \{
		\x \in \GG_N \} \quad \text{and} \quad w_2(\x) \equiv
		\{ \x \in \GG_N^\mathsf{c} \}.
	\end{equation} 
	Here, $\GG_N$ is an estimate of the
	failure region in \eqref{eq:failure_region}, say via GP $\mathcal{S}_N$
	predictive equations in
	\eqref{eq:pred_eqs}.  We shall be more precise momentarily.  Abusing notation
	slightly, this $H(W)$  would highly value inputs $\x$, via larger
	entropy, which are close to $\CC_N$, the (estimated) boundary of
	$\GG_N$, i.e., such that $\mathbb{P}(w_1) \approx
	\mathbb{P}(w_2) \approx 1/2$, which is sensible.
	
	\cite{oakley2004estimating} considered entropy for choosing among candidate
	LHSs in the second step of a two-stage design setup, whereas we consider 
		a pure-sequental setup where the number of stages is proportional to the desired design size.
	They targeted output quantiles, like
	95\%, implicitly defining $\CC$ and thus $\GG$ and their GP-estimated analogues.
	We are interested in much larger quantiles, such that any LHS candidate would
	have low probability of being nearby $\GG_N$, mirroring a calculation similar
	to that in Eq.~$\eqref{eq:monte_carlo}$.
	Contour
	Location Via Entropy Reduction \citep[CLoVER;][]{marques2018contour} greedily
	maximizes the reduction in contour entropy through a lookahead scheme. As in
	several other methods, CLoVER relies on a set of reference points or knots in
	$\mathcal{X}$ for numerical integration.  But knots are problematic for reasons
	similar to the LHS candidates above: hard to get enough nearby/inside the
	failure region.  More knots do not help much because the expense of quadrature
	is squared. 
	\section{Entropy-based Adaptive Design}
	\label{sec:entropy}
	MFIS holds the potential to leverage a cheaper surrogate to obtain
	unbiased reliability estimates, but only if the surrogate is accurate around $\CC$.   Here we propose an entropy-based contour locator (ECL)
	acquisition function toward that goal.
	
	\subsection{Entropy acquisition function}
	\label{ss:acquisition}
	
	After being fit to training data $D_N$, our GP surrogate $\mathcal{S}_N$
	provides a distribution in \eqref{eq:pred_eqs} for $Y(\x)$. In the case of a
	failure region $\GG$ defined by affine limit state function $g$, the entropy 
	\eqref{eq:gen_entropy} with discrete events \eqref{eq:ecl_categories} for input $\x$ relative to estimate $\GG_N$ via
	$\mathcal{S}_N$ is
	\begin{align}
		\mathrm{ECL}&(\x\mid\mathcal{S}_N,g)=  \label{eq:our_entropy} \\
		&- \mathbb{P}\Big(g(Y(\x))> 0\Big)\log \mathbb{P}\Big(g(Y(\x))> 0\Big) 
		- \mathbb{P}\Big(g(Y(\x))\leq0\Big)\log \mathbb{P}\Big(g(Y(\x))\leq0)\Big) \nonumber \\
		&=-\left(1-\Phi\left(\frac{\mu_N(\x)-T}{\sigma_N(\x)}\right)\right)\log\left(1-\Phi\left(\frac{\mu_N(\x)-T}{\sigma_N(\x)}\right)\right) \nonumber \\
		&\qquad \qquad -\Phi\left(\frac{\mu_N(\x)-T}{\sigma_N(\x)}\right)\log\left(\Phi\left(\frac{\mu_N(\x)-T}{\sigma_N(\x)}\right)\right), \nonumber 
	\end{align}
	where $\Phi$ is the standard univariate Gaussian CDF.  Noting the symmetry in
	\eqref{eq:our_entropy}, this would work identically for any $g(y)=a(y-T)$
	where $a
	\in \{-1,1\}$, meaning that the underlying criterion is unchanged when
	searching for extremely small, rather than large, thresholds. Also,
	ECL trends higher as the GP's predicted mean $\mu_N(\x)$ moves closer to the
	threshold or $\sigma_N(\x)$ is higher, facilitating an
	exploration--exploitation trade off.  ECL is maximized when
	$\mathbb{P}(g(Y(\x))> 0)=\mathbb{P}(g(Y(\x))\leq0)=1/2$, targeting $\x \in
	\CC_N \equiv \{\x: g(\mu_N(\x))=0\}$.
	
	\begin{figure*}[ht!]
		\includegraphics[trim=0 5 0 35,clip, width=.6\textwidth]{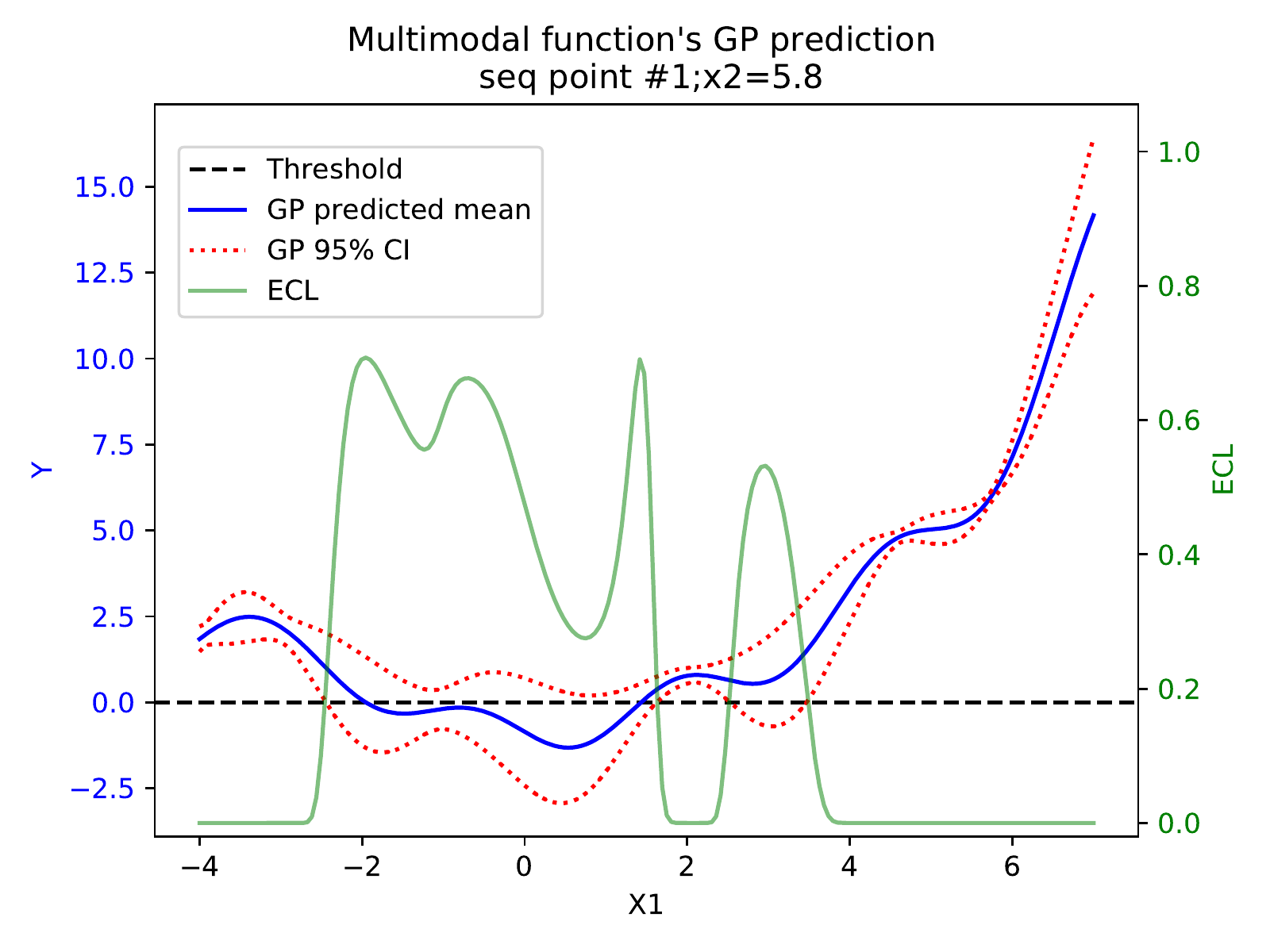}
		\centering
		\caption{GP surrogate predictions and ECL values for the multimodal function along $x_2=5.8$.}
		\label{fig:entropy_slice}
	\end{figure*}
	
	To explore the ECL surface, consider a 2d multimodal function
	\citep{bichon2008efficient}:
	\begin{equation}\label{eq:mm_function}
		\mathcal{T}^{\text{MM}}(\x)=\frac{(x_1^2+4)(x_2-1)}{20}-\sin\left(\frac{5x_1}{2}\right)-2,
	\end{equation}
	where $x_1 \in [-4,7], x_2 \in [-3,8]$. This function has been used for
	reliability/level-set finding with $T=0$ \citep{bichon2008efficient, marques2018contour}. We trained a GP on an LHS
	of $N=20$ points.  Figure \ref{fig:entropy_slice} shows the predictive mean
	($\mu_N(\x)$ blue line), 95\% mean confidence interval (roughly $\pm 2
	\sigma_N(\x)$, red dotted lines) and ECL (green line) across the slice $x_2=5.8$.
	Since $g(\mu_N(\x))$ crosses zero twice, there are two global maxima of ECL in
	this slice alone, along with several other local maxima where the upper bound of the confidence bound is close to or above zero. In higher volume input
	spaces the dimension of the manifold tracing out the level set also increases,
	resulting in a continuum of ridges in the ECL surface.  If you want to look
	ahead, Figure \ref{fig:ent_surface} provides a visual. Such a surface
	could present challenges to effective numerical optimization for the purpose
	of solving for adaptive design acquisitions.
	
	\subsection{Optimization-based acquisition}
	\label{ss:opt}
	
	To address the (potentially) multimodal nature of ECL in \eqref{eq:gen_entropy}, as illustrated in Figure \ref{fig:entropy_slice}, we developed a two-step
	strategy to solving for each greedy acquisition:
	\[
	\x_{N+1} = \mathrm{argmax}_{\x \in \mathcal{X}} \mathrm{ECL} (\x \mid \mathcal{S}_N).
	\]
	In the first stage we optimize over a discrete-space-filling candidate set
	$\bar{\X}_{N_c}$.  We prefer LHS candidates with $N_c \approx 10d$,
	following the rule of thumb proposed in \citet{loeppky:etal:2009}.\footnote{This paper
		targets design sizes for computer experiments, not numbers of active learning
		candidates, but we find the logic of the two to be quite similar.}  Then, in
	the second stage, we use this global, but coarse, solution to initialize a
	local solver, such as BFGS \citep{byrd1995limited}.  The first stage is
	designed to ``select'' a global domain of attraction, while the second stage
	ascends that peak of the ECL surface.
	
	\begin{algorithm} \caption{Entropy Optimization} 
		\begin{algorithmic}[1] 
			\Procedure
			{Entropy.Opt}{$N_c, \mathcal{S}_N, \mathcal{X}$}
			\State $\bar{\X}_{N_c} \leftarrow$ LHS of size $N_c$ in $\mathcal{X}$ \hfill \Comment{Candidate set}
			\State $\check{\x} \leftarrow \mathrm{argmax}_{\bar{\x}\in \bar{\X}_{N_c}} \; \mathrm{ECL}(\bar{\x} \mid \mathcal{S}_N)$ \hfill \Comment{Discrete search}
			\State $\x_{N+1} \leftarrow \mathrm{argmax}_{\x \in \mathcal{B}(\check{x})} \; \mathrm{ECL}(\x \mid \mathcal{S}_N)$ \hfill \Comment{Continuous, local, from $\check{\x}$}
			\State \Return $\x_{N+1}$
			\EndProcedure 
		\end{algorithmic} \label{alg:entropy_opt} 
	\end{algorithm}
	
	The details are laid out in pseudocode in  Algorithm \ref{alg:entropy_opt}. In
	line 4, representing stage 2 local search, the scope is based on $\check{\x}$.
	For a true local search, the region $\mathcal{B}(\check{\x})$ may comprise of
	a box containing $\check{\x}$ whose sides extend to nearby elements of the
	candidate set $\bar{\X}_{N_c}$ or to the edge of $\mathcal{X}$, whichever is
	closer.  Such narrowed scope may help local optimizers that support box
	constraints, like BFGS does. In our implementation (more details in Section
	\ref{ss:implement}) simply take $\mathcal{B}(\check{\x}) = \mathcal{X}$, the
	whole space, and use $\check{\x}$ to initialize the search.
	
	We find that the stochastic and
	coarse global nature of the first stage (owing to the small LHS), paired with
	a precise and focused second stage, brings benefits beyond avoiding solutions
	trapped in vastly inferior local maxima.  Sometimes the surrogate
	$\mathcal{S}_N$ becomes over-confident about its predictive equations, leading
	to small $\hat{\sigma}(\x)$ nearby an estimated failure region contour
	$\mathcal{C}_N$ that is far from the true $\mathcal{C}$. In such situations,
	the first stage offers potential to ignore that area for local maxima in a less seductive, but
	almost-as-good region.  Injecting a degree of stochasticity into
	optimization is a convenient way of avoiding pathologies, building-in
	robustness \citep{wolpert1997no}.
	
	\begin{figure*}[ht!]
		\centering
		\includegraphics[trim=0 0 0 0,clip, width=\textwidth]{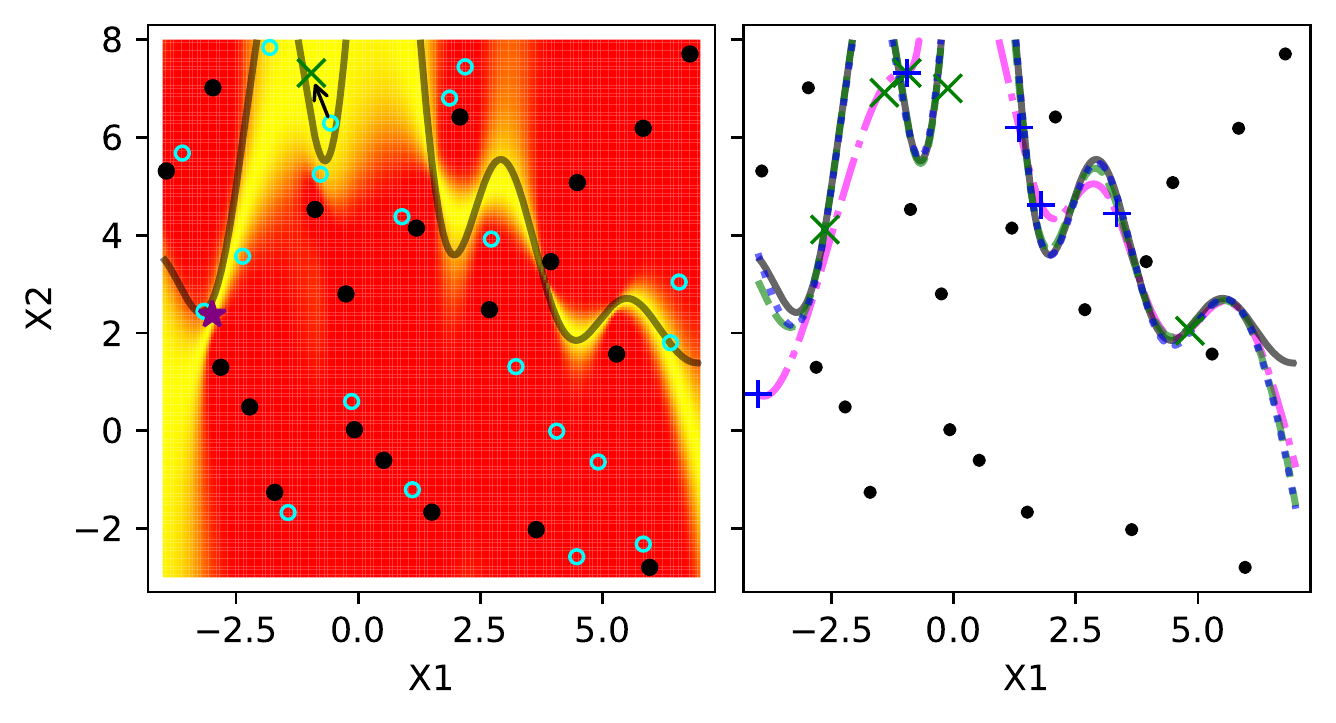}
		\caption{{\em Left:} ECL surface based on a GP fit to $N_0=20$ points (black circles). The true $\CC^{\text{MM}}$ is denoted by the black curve. Candidates $\X_{N_C}$ are aqua open circles, 
			with $\check{\x}$ at the terminus of an arrow pointing to $\x_{N_0+1}$, the optimized selection as green `$\times$'. The purple star shows the best of these from all candidates. {\em Right:} zero contours based on initial GP fit ($\mathcal{S}_{N_0}$ pink dot-dashed curve), at $\mathcal{S}_{N_0+5}$, after five ECL sequentially selected points (green $\times$'s/dashed), and similarly after a single $N_b=5$ batch $\mathcal{S}_{N_0+N_b}$ (blue pluses/dotted curve).}
		\label{fig:ent_surface}
	\end{figure*}
	
	An illustration on the 2d multimodal function \eqref{eq:mm_function} is
	provided in the left panel of Figure \ref{fig:ent_surface}. Here we show the
	ECL surface based on a GP $\mathcal{S}_N$ fit to the same $N_0=20$ points as
	used for Figure \ref{fig:entropy_slice}. The multimodal function's true zero contour 
	($\CC^{\text{MM}}$) is shown with the black curve. Observe that the highest ECL
	regions (yellow) are near that contour. Other local maxima are present where
	there is little data (bottom left corner) and $\mathcal{S}_N$'s uncertainty
	high. Aqua blue open circles indicate the candidate set $\bar{\X}_{20}$. The arrow
	pointing from one blue circle to the green `$\times$' denotes the chosen
	candidate point $\check{\x}$ from the first stage and the resulting $\x_{N+1}$
	after continuous optimization in the second. Notice that the green `$\times$' is different from the purple star, the optimal location if local optimization is conducted on all 20 candidates. The green `$\times$' finds an  almost-as-good region that is close to $\CC$.
	
	Although technically (slightly) sub-optimal for the individual acquisition of
	$\x_{N+1}$, we think that the view in Figure \ref{fig:entropy_slice} (left)
	suggests that the solution we found (green `$\times$') is indeed sensible. The
	first stage is more likely to target a high uncertainty region, with a wider
	domain of attraction (more yellow), due to its coarse nature.  Ideally, the
	attractiveness of ``big yellow'' regions would be captured formally by the
	acquisition criteria.  This is why most authors integrate over the input
	space, but that requires numerical quadrature which we are trying to avoid.
	Our two-stage scheme is simpler to implement, and as results shown later in Section \ref{ss:syn_results} support, fast in execution and robust in
	a loose sense.

	\subsection{Batch selection}
	\label{ss:batch}
	
	Modern high-performance computing (HPC) resources allow simulations to be run
	in parallel. To accommodate, here we extend our ECL adaptive design strategy to
	batch acquisition. Suppose we wish select a batch of $N_b$ samples, e.g., to
	saturate the node of a computing cluster. One option is to extend our criteria in
	\eqref{eq:our_entropy} from singular $\x$ to multiple $\X_{N_b}$, in the
	spirit of \citet{zhang2020batch} for variance-based acquisition.  This is
	doable by upgrading surrogate $\mathcal{S}_N$ pointwise predictive
	equations in \eqref{eq:pred_eqs} to joint ones. Similar MVN
	conditioning applies, e.g., see Eq.~(5.3) in \citet{gramacy2020surrogates}.
	Next, calculate entropy in \eqref{eq:gen_entropy} on the resulting MVN.  You will
	still get a closed form (not shown) up to multivariate CDF evaluations
	$\Phi_{N_b}$. Calculating $\Phi_{N_b}$ can pose a challenge.  Although univariate $\Phi$ involve a
	degree of quadrature, evaluation is so fast and accurate (and built-in as
	native in most programming languages) that we generally regard it as analytic
	in practice. Multivariate $\Phi_{N_b}$, involving high dimensional quadrature
	in vast ``tail volumes'', is much more challenging, see \cite{genz2009}.  Although software is available for {\sf R}
	\citep{mvtnorm}, we find that $N_b$ on the scale of the number of cores in
	supercomputers (e.g., $N_b=16$ or bigger), evaluation is slow and with
	accuracy that can be problematic for downstream library-based optimization via methods, e.g., via BFGS.
	
	Instead we prefer to select batch elements sequentially, updating
	$\mathcal{S}_N \rightarrow \mathcal{S}_N^{(+1)}$ to account for each new
	selected input (but not its response) in the batch until we reach
	$\mathcal{S}_{N}^{(+N_b)}$.  Specifically, given $0 \leq n_b \leq N_b$
	selections so far, we may deduce the following equations for
	$\mathcal{S}_{N}^{(+n_b)}$, where $\X_{N+n_b}=\X_N \cup \X_{n_b}$
	and with $\X_{0} \equiv \emptyset$ reducing to the following using \eqref{eq:pred_eqs}:
	\begin{align} \label{eq:gp_update_pred_eqs}
		\mu_{N}^{(+n_b)}(\x'|\X_{N+n_b},\Y_N) &= \mu_{N} (\x'|\X_{N},\Y_N) =\kk(\x', \X_N)\kk(\X_N, \X_N)^{-1}\Y_N \\
		\sigma^{2(+n_b)}_N(\x'|\X_{N+n_b},\Y_N)&= k_{\Psi}(\x',\x')-\kk(\x', \X_{N+n_b})\kk(\X_{N+n_b}, \X_{N+n_b})^{-1}\kk(\x', \X_{N+n_b})^\top. \nonumber
	\end{align} 
	Each of these updates $\mathcal{S}_{N}^{(+n_b)} \rightarrow \mathcal{S}_N^{(+n_b+1)}(\x_{n_b+1})$, in particular for $\sigma^{2(+n_b)}(\x)$ as $n_b
	\rightarrow n_b+1$ with new acquisition $\x_{n_b+1}$ augmenting $\X_{n_b}$ to
	build $\X_{n_b+1}$, can be performed in time quadratic in $N+n_b$ via partition
	inverse equations \citep[][Section
	6.3]{barnett:1979,gramacy2011particle,gramacy2020surrogates}.  However, the
	otherwise cubic burden is often manageable in active learning where the whole
	enterprise is focused on making $N$ as small as possible.
	
	\begin{algorithm} \caption{Batch Entropy} 
		\begin{algorithmic}[1] 
			\Procedure
			{Entropy.Batch}{$\mathcal{S}_N$, $N_b$, $N_c$, $\mathcal{X}$}
			
			\State ${\mathcal{S}}_{N}^{(+0)} \leftarrow \mathcal{S}_{N}$ \hfill \Comment{Initialize deduced surrogate}
			\For{$n_b=1,\dots,N_b$}
			\State $\x_{n_b} \leftarrow \text{\sc Entropy.Opt}(\mathcal{S}_{N}^{(+n_b-1)}, N_c, \mathcal{X})$ \hfill \Comment{Select next point (Alg.~\ref{alg:entropy_opt})}
			\State $\mathcal{S}_{N}^{(+n_b)} \leftarrow \mathcal{S}_{N}^{(+n_b-1)} (\x_{n_b})$ \hfill \Comment{Update variance Eq.~\eqref{eq:gp_update_pred_eqs} in GP}
			\EndFor
			\State \Return ${\X}_{N_b} \equiv [\x_1^\top, \dots, \x_{N_b}^\top]^\top$ \hfill \Comment{Return batch of $N_b$ design points}
			\EndProcedure 
		\end{algorithmic} \label{alg:batch_entropy} 
	\end{algorithm}
	
	With $\mathcal{S}_{N}^{(+N_b)}$ defined in this way, batch selection becomes
	an exercise in repeated greedy selection.  This is detailed in Algorithm
	\ref{alg:batch_entropy}, with Algorithm \ref{alg:entropy_opt} deployed as a
	subroutine. It is worth remarking that a new candidate set $\bar{\X}_{N_c}$ is
	generated for each new call of line 4. If we were to rely upon the same
	candidate set for the entire batch, we would in effect be taking the $N_b$
	best candidate points, some of which could be far from the local maxima
	regions of interest. It is possible (although in practice uncommon) that some
	of the acquisitions are very close to others in the same batch. Since we
	operate under the assumption that the high-fidelity model is deterministic,
	replicates are not helpful. In our implementation (more in Section
	\ref{ss:implement}), we account for this by reverting to the optimal candidate
	point $\check{\x}$ for $\x_{n_b}$ in such cases, undoing the stage 2 local
	search.  At the end of the batch, new simulations must be performed at
	$\X_{N_b}$, forming $\Y_{N_b}$, combining with $D_N$ to obtain $D_{N+N_b}$.
	Inference follows for any hyperparameters $\Psi$ required to form
	$\mathcal{S}_{N+N_b}$ which is, of course, different than the intermediate
	$S_{N}^{(+n_b)}$ which may be discarded after each iteration of the loop.
	
	To illustrate, we return to the multimodal example of Section \ref{ss:opt}.
	The right panel of Figure \ref{fig:ent_surface} compares the acquisition of
	five runs one-at-a-time (i.e., multiple single acquisitions with updates) and
	five as a batch.  The dash-dotted pink line shows the GP's predicted mean
	contour via $\mathcal{S}_{N_0}$ before either adaptive design. The green
	$\times$'s were selected one at a time, leading to the green dashed
	$\mathcal{C}_{N+5}$ contour. In a similar fashion, the blue $+$'s and dotted
	curve represent the results from the batch size of 5. Both adaptive designs
	select the same first point (close to (-1.25, 7.2)); this is guaranteed by using the same candidate set. In the
	batch design, only the GP's variance equation is updated, which results in
	runs being selected close to the original GP's zero contour were ECL is
	maximized. In contrast, repeated one-at-a-time acquisition allows updating of
	the mean surface. Yet surprisingly, the resulting contours from both designs
	(green dashed and blue dotted lines) are fairly similar and capture
	$\CC^{\text{MM}}$ well.  This mirrors other work on batch active
	learning with GPs \citep{zhang2020batch}; little is lost on the batch setting
	versus its purely greedy analog.
	
	\section{Implementation and benchmarking}
	\label{sec:toy_results}
	
	Now we provide results for a series of experiments showcasing ECL, and
	contrasting against existing contour-finding acquisition functions. We first
	detail our implementation and the synthetic simulators used in the
	experiments. Then we present and discuss results of MC sequential design
	exercises in terms of sensitivity and relative error of the failure region's
	volume based on GP predictions in \eqref{eq:pred_eqs} over acquisition
	iterations.
	
	All analysis was performed on a six-core hyperthreaded Intel i7-9850H CPU at
	2.59GHz. {\sf Python} and {\sf R} code \citep{R} supporting our methodological
	contribution, and all examples reported here and throughout the paper, may be
	found on our Git repository.
	\begin{center}
		\url{https://bitbucket.org/gramacylab/nasa/src/master/entropy}
	\end{center}
	
	\subsection{Implementation and synthetic data}
	\label{ss:implement}
	
	The following implementation details are noteworthy. Our methodological
	contribution is encapsulated in the {\sf Python} package {\tt eclGP},
	incorporating GP functionality from the {\tt sklearn} package
	\citep{scikit-learn}. For all of our synthetic examples, we privilege a
	separable Gaussian kernel formulation, although other forms such as Mat\'ern
	\citep{Stein2012} could easily be used. To ensure stability in the calculation
	of the inverse of the covariance matrix, we fix a jitter \citep{Neal1998} of
	$10^{-6}$.
	
	We compare the ECL-based estimates of a contour $\mathcal{C}$ to
	established contour finding acquisition functions with publicly available code. For CLoVER, we utilize the
	{\sf Python} code accompanying \cite{marques2018contour}, with the only
	adjustment being setting the candidate and integration knots to LHSs
	of size 1000 for each experiment (see Appendix \ref{app:grid_density} for a discussion). Implementations for all other methods
	leverage the {\sf R} package {\tt KrigInv} \citep{chevalier2014fast}, using
	the \verb!"genoud"! option for optimization and the recommended defaults for
	constructing integration knots where relevant (SUR, tIMSE). In  {\sf R}, we
	rely on {\tt DiceKriging} \citep{dicekriging} to fit GPs.

	\begin{table}[ht!]
		\centering
		\begin{tabular}{l|c|r|r|c|r|r} 
			\hline
			Function &  $\mathcal{X}$ & $N_0$ & $N$ & $g(y)$ & vol($\mathcal{G}$) & Quantile \\ 
			\hline
			Branin-Hoo &  $[-5,10]\times[0,15]$ & 10 & 30 & $y-206$ & 2.1783 & 0.9903\\ \hline
			Ishigami &  $[-\pi,\pi]^3$ &30 & 200 & $-y+10.244$ & 0.0250 & 0.9999\\ \hline
			Hartmann-6 &  $[0,1]^6$ & 60 & 500 & $y-2.63$ & 0.0011 &0.9989
		\end{tabular}
		\caption{The design space, allocation of samples, limit state functions, volume of the failure region, and quantiles for the synthetic tests.\label{tab:benchmark_details}}
	\end{table}
	
	Experiments are conducted on the Branin-Hoo \citep{forrester2008engineering},
	Ishigami \citep{ishigami1990importance}, and Hartmann-6
	\citep{surjanovicvirtual}\footnote{following
		\url{https://www.sfu.ca/~ssurjano/hart6.html}} functions. Table
	\ref{tab:benchmark_details} provides a summary of the experiments, including
	the number of sample points $N$ and limit state functions $g(\cdot)$ used to
	define failure regions $\mathcal{G}$. The limit state functions are defined so
	that $\mathcal{G}$ for each function contains at least two local extremes
	located in disjoint failure regions. The volume of $\GG$ is less than 1\% of
	the total volume of the design region in each instance. Volumes reported in
	the table are based on averaging 100 MC estimates from LHSs of size $10^6$.
	Dense testing LHSs were used to calculate sensitivity and predicted volume of
	the failure region out-of-sample. For Branin-Hoo an LHS of $5\times 10^6$
	points is used for prediction/classification, whereas Ishigami and Hartmann-6
	used $10^7$.  To initialize the sequential design we follow the rule of thumb
	for setting $N_0, N_c=10d$ \citep{loeppky:etal:2009}, except for the 2d
	Branin-Hoo where $N_0=10$ was sufficient.

	\subsection{Synthetic tests for contour finding}
	\label{ss:syn_results}
	
\begin{figure*}[]
		\centering
		\includegraphics[trim=0 0 5 15, clip,width=\textwidth]{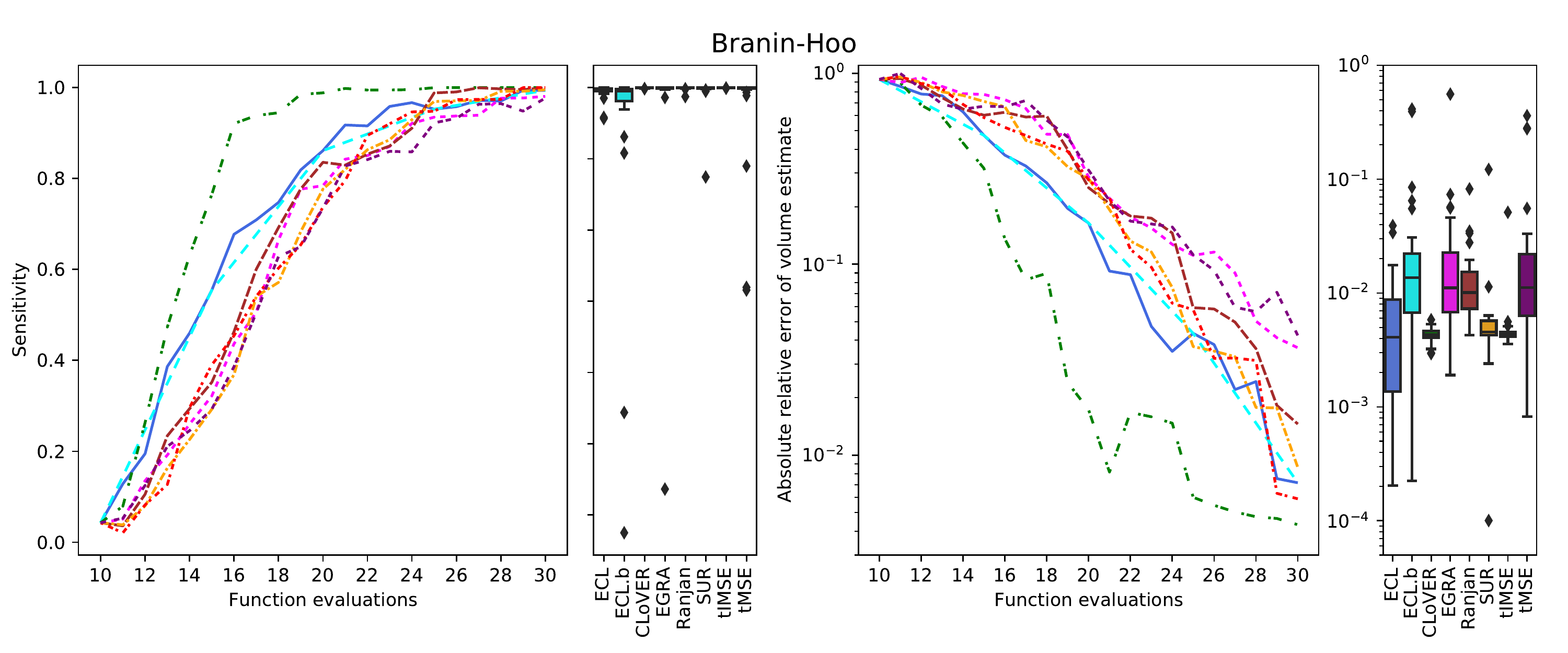}
		\includegraphics[trim=0 0 5 15, clip,width=\textwidth]{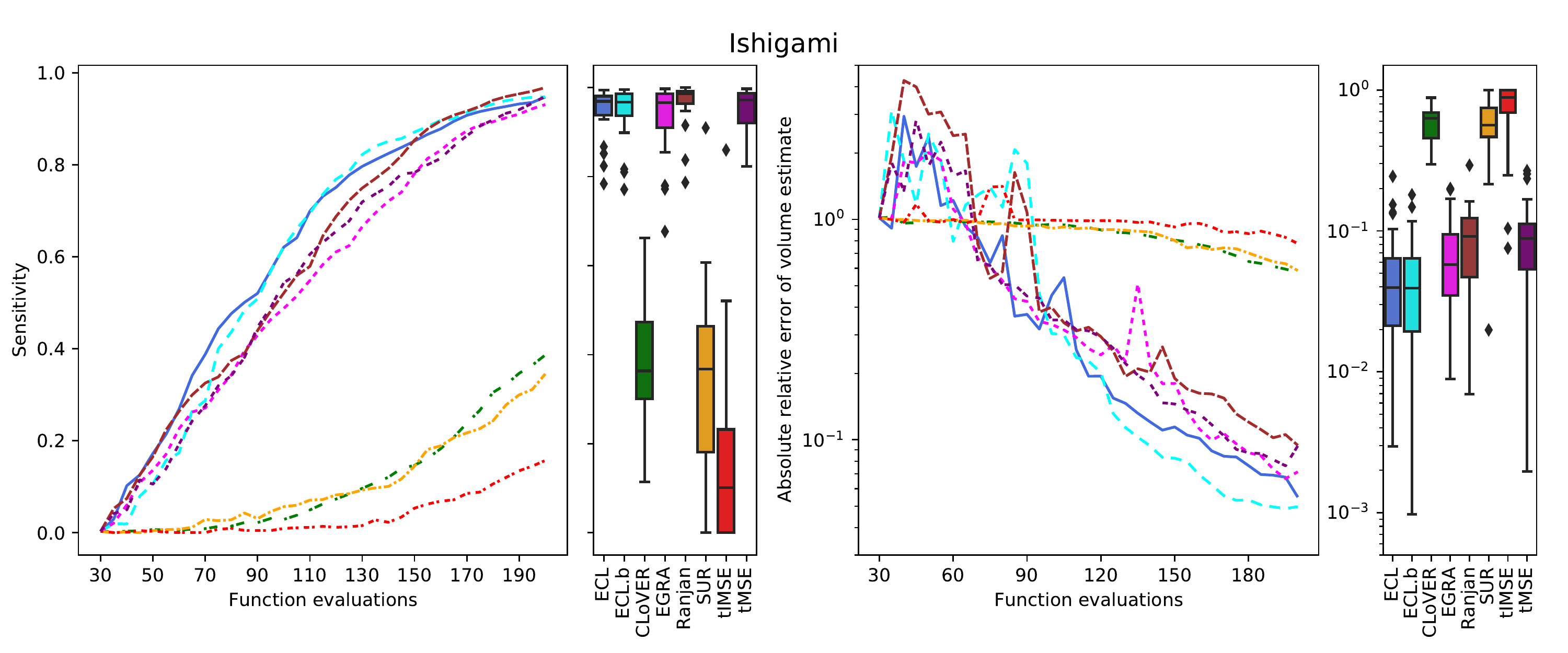}
		\includegraphics[trim=0 0 5 15, clip,width=\textwidth]{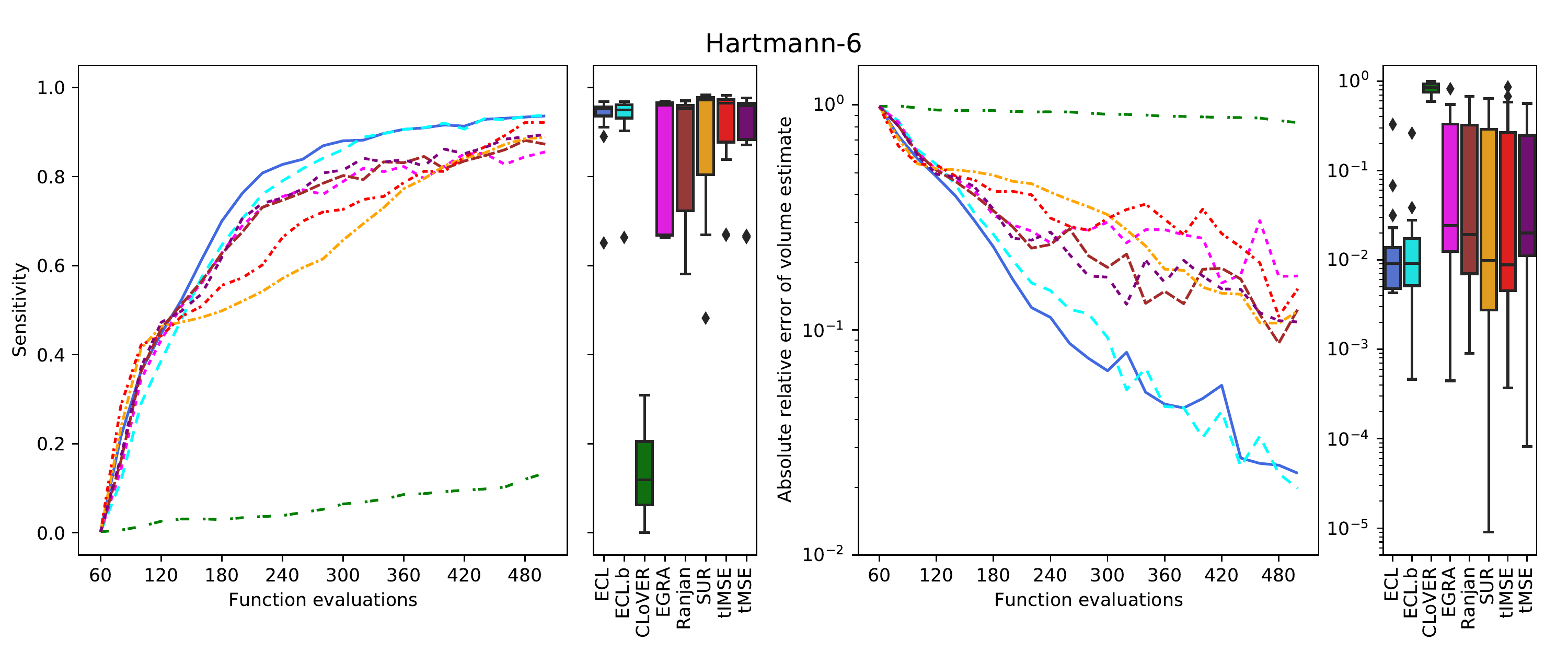}
		\includegraphics[trim=50 18 200 320, clip,width=\textwidth]{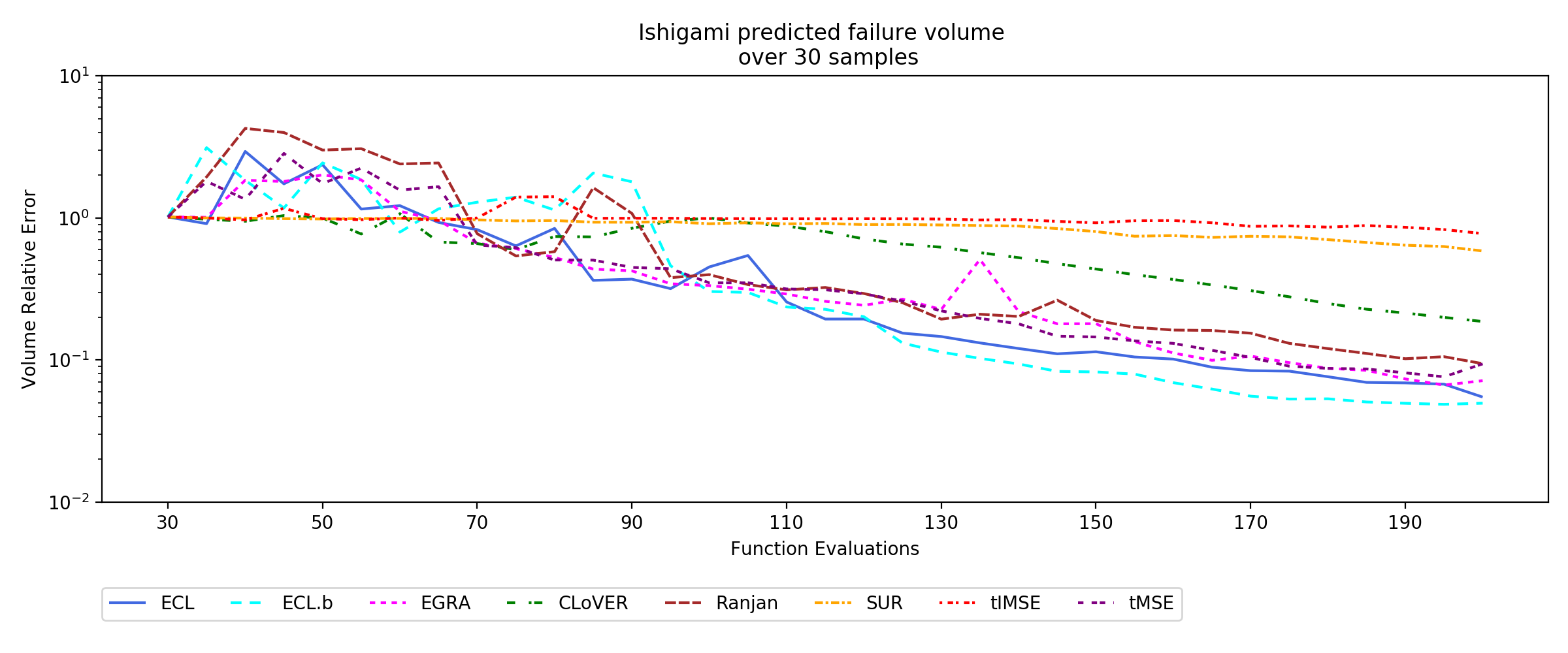}
		\caption{Comparing sequential designs in 30 MC repetitions. {\em Left:} mean
			sensitivity  classifying failure region $\mathcal{G}$. {\em Middle left:}
			distributions of sensitivity after last acquisition. {\em Middle right:}
			mean relative error $\mathrm{vol}(\mathcal{G})$. {\em Right:} final
			distributions of relative volume error. }
		\label{fig:benchmark_plots}
	\end{figure*}

	Earlier in Figure \ref{fig:ad_vs_sf_bakeoff}, when motivating adaptive design
	strategies for contour location, we showcased results on the Ishigami
	function. Those showed that using an ECL design of 200 points provides a GP
	predictive mean that produces much higher sensitivity and specificity than a
	GP trained on an LHS of the same size. Here we expand that analysis against
	benchmarks.  Figure \ref{fig:benchmark_plots} augments with other test
	functions and the following competitors: EGRA \citep{bichon2008efficient},
	Ranjan \citep{ranjan2008sequential}, and tMSE \citep{picheny2010adaptive} as
	strategies that use closed-form calculations for continuous optimization
	acquisition; CLoVER \citep{marques2018contour},  SUR \citep{chevalier2014fast}
	and tIMSE \citep{picheny2010adaptive} require integral approximation.
	
	The figure tracks sensitivity as design size increases (left) and provides a
	boxplot of the final distribution of sensitivity (middle left). Similarly,
	relative error for  $\mathrm{vol}(\hat{\GG})$ is tracked (middle right),
	followed by its final distribution error (right). Thirty MC repetitions were
	conducted for each experiment, with the line plots displaying means.
	Across all three experiments (rows in the figure) our ECL method is among the
	best achievers in terms of sensitivity (higher is better) and volume 
	error (lower is better).
	
	For Branin-Hoo, ECL and CLoVER -- the entropy-based methods -- distinguish
	themselves early on in the adaptive design in both sensitivity and volume
	error compared to all other methods. By the end of the experiment, all methods
	have a similar sensitivity. ECL's volume error is in line with the integration methods, beating its closed-form
	competitors.  For Ishigami, observe that final mean sensitivity for ECL is
	bested by Ranjan. Yet when it comes to the relative error of the contour's
	volume, ECL is far below that of Ranjan.  Hartmann-6 imposes challenges in
	higher input dimension, which is why we extend the adaptive design to 500
	total points. In this experiment, we find that the explorative nature of ECL
	is key to finding the failure contour. The bottom row of Figure
	\ref{fig:benchmark_plots} shows that ECL provides a much more robust design
	with high sensitivity and a volume error nearly an order of magnitude
	smaller than others. We also created a batch ECL design (ECL.b) with batch sizes of 5 for Branin-Hoo and 10 for Ishigami and Hartmann-6. In all three experiments, ECL.b performs similarly
	to ECL.
	
	For the methods that rely upon a ``grid'' of integration knots (CLoVER, SUR,
	tIMSE), the size of that grid greatly affects their ability to explore and
	identify parts of the failure contour. With a set of 1000 knots for
	Branin-Hoo, CLoVER performs very well -- reaching a high sensitivity plateau
	before the other methods. Yet when the same number of knots is used in the 3d
	or 6d problems with a higher quantile, it struggles. In a similar way, the
	combination of multiple failure regions (six) and a high quantile for the
	Ishigami experiment poses a great challenge for SUR and tIMSE. See Appendix
	\ref{app:grid_density} for further discussion.
	
	\begin{table}[ht!]
		\centering
		\begin{tabular}{l|r|r|r|r|r|r|r|r} 
			\hline
			Method & ECL & ECL.b & CLoVER & EGRA  & Ranjan & SUR & tIMSE & tMSE \\ 
			\hline
			\textbf{Branin-Hoo}& 0.008 & 0.002 & 0.42 & 0.2  &  0.2 & 0.4 & 0.3 &  0.2\\ \hline
			\textbf{Ishigami}& 0.10  & 0.08 & 17.8 & 3.9 &  4.5 & 3.5 & 3.4 &  3.9 \\ \hline
			\textbf{Hartmann-6}& 4.40 & 0.78 & 428.4 & 66.0 & 59.2 & 109.1 & 111.7 & 59.5 
		\end{tabular}
		\caption{Average computation times (minutes) to select points across MC 30 repetitions.}
		\label{tab:timings}
	\end{table}

	Table \ref{tab:benchmark_details} summarizes average computation time to build
	one full sequential design for each method. The timings include acquisition
	efforts and subsequent GP updating. Scripts for ECL, ECL.b and CLoVER are in
	{\sf Python} and the rest are in {\sf R}. Note the speed at which ECL and
	ECL.b adaptive designs are built, especially compared to methods using
	numerical quadrature (CLoVER, SUR, and tIMSE). ECL provides between one and
	two orders of magnitude speedup. Using a batch size of 10 cuts the average
	computation time of ECL by five for the Hartmann-6 experiment, while still
	providing designs with similar sensitivity and volume error.

	\section{Failure probability estimation}
	\label{sec:xemu_results}
	
	Here we transition from contour finding to failure probability estimation,
	combining our ECL adaptive design with MFIS as in Figure \ref{fig:flowchart}.
	We refer back to the synthetic functions of Section \ref{sec:toy_results}
	before applying the methodology to our motivating spacesuit impact simulator.
	
	\subsection{Synthetic benchmarking}
	
	Figure \ref{fig:benchmark_plots} illustrates how sequential design can furnish
	accurate GP mean predictions (high sensitivity and low relative error) in the
	vicinity of a failure region. As introduced in Section \ref{ss:failure_prob}
	and foreshadowed in Figure \ref{fig:ad_vs_sf_bakeoff}, an unbiased estimate of
	this volume (i.e., the failure probability) may be improved through further
	sampling with MFIS. Here we build upon Section \ref{sec:toy_results}
	experiments with the Ishigami and Hartmann-6 functions. With GPs
	$\mathcal{S}^{\text{Ish}}$ and $\mathcal{S}^{\text{Hart}}$ trained to the ECL
	adaptive designs as surrogates, or ordinary LHSs as a
	benchmark, we use our {\sf Python} package {\tt
		MFISPy}\footnote{https://github.com/nasa/MFISPy} to generate our bias distributions and calculate $\hat{\alpha}$.
	The {\tt scipy.stats} package was used for the input distributions.
	
	Ishigami used input distribution $\F^{\text{Ish}}$ with independent marginals  $x_1\sim
	\mathcal{N}(-1, 1)$, $x_2\sim \mathcal{N}(1.5,
	1.5^2)$ and $x_3 \sim \text{Uniform}(-\pi, \pi)$, where the Gaussians are
	truncated to $\mathcal{X} \in [-\pi,\pi]^3$, which puts $\F^{\text{Ish}}$'s
	center of the mass near to four of the six failure regions.\footnote{A modification to the original experiment in Section \ref{ss:adaptive_design} (where uniform distributions were used).} Combining the
	limit state function in Table \ref{tab:benchmark_details} along with this
	input distribution results in $\alpha^{\text{Ish}}\approx1.9\times10^{-4}$.
	For Hartmann-6 we used $\F^{\text{Hart}}$ comprised of independent $x_j \sim
	\mathcal{N}(0.5, 0.1^2)$ for $j=1,\dots,6$ truncated to $\mathcal{X} =
	[0,1]^6$. Here the failure probability, using the threshold in Table
	\ref{tab:benchmark_details}, is
	$\alpha^{\text{Hart}}\approx9.96\times10^{-6}$.
	
	Following Algorithm \ref{alg:importance_sampling}, we generated
	$M^{\text{Ish}}=5\times10^6$ samples from $\F^{\text{Ish}}$, obtaining
	predictive evaluations under $\mathcal{S}^{\text{Ish}}$ at those locations. Due to
	$\alpha^{\text{Hart}} \ll  \alpha^{\text{Ish}}$, we produced many more
	$M^{\text{Hart}}=5\times10^7$ samples from $\mathcal{S}^{\text{Hart}}$ in
	order to obtain enough failures for the bias distribution training,
	$\F_\star^{\text{Hart}}$. We fit each bias
	distribution using a Gaussian mixture model of up to 10 clusters, determined
	with cross-validation, using diagonal covariances. Both experiments operated
	with a total budget of 1000 evaluations of the true model $\mathcal{T}$. Thus,
	based on the adaptive design budgets were $N=200$ for Ishigami and 500 for
	Hartmann-6, $M^\star$ = 800 and 500 respectively.
	
	In many engineering applications, budget considerations limit data collection
	to a single experiment. Generating a single estimate $\hat{\alpha}$ motivates
	a desire to be conservative in the estimate
	\citep{auffray2014bounding,azzimonti2020adaptive}. As discussed by
	\cite{dubourg2011metamodel}, we can leverage the GP's posterior distribution
	when classifying failure inputs used to fit the bias distribution. Namely, we
	define $\hat{y}(\x)$ in line 4 of Algorithm \ref{alg:importance_sampling} as
	$\hat{y}(\x)=\mu_N(\x)+\delta \sigma_N(\x)$, with $\mu_N(\x),\sigma_N(\x)$
	from \eqref{eq:pred_eqs} . For our problems we use $\delta=1.645$, making
	$\hat{y}(\x)$ the 95\% upper confidence bound (UCB).
	
\begin{figure*}[ht!]
		\centering
		\includegraphics[trim=20 0 0 18, clip,width=0.51\textwidth]{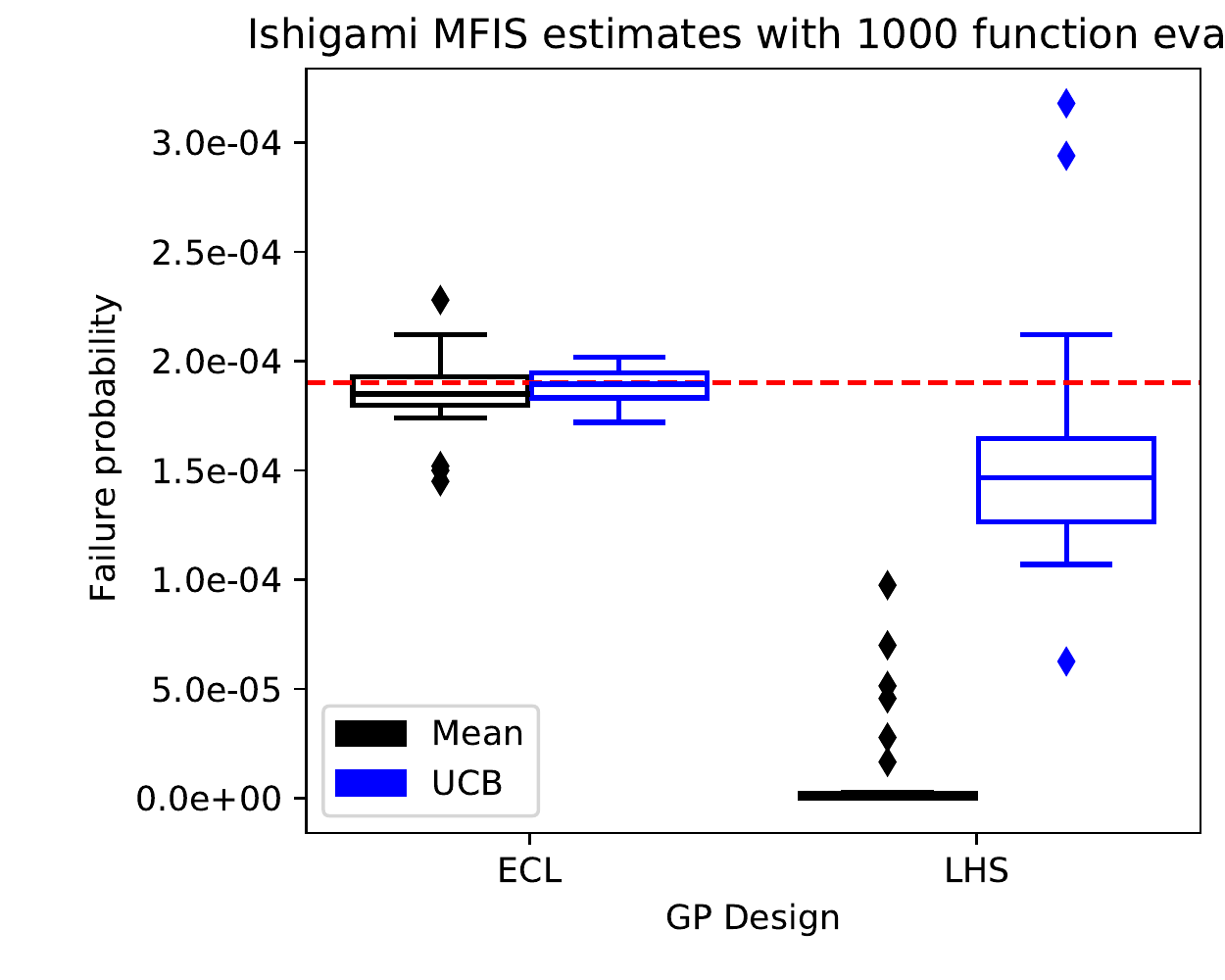} 
		\includegraphics[trim=40 0 0 18, clip,width=0.48\textwidth]{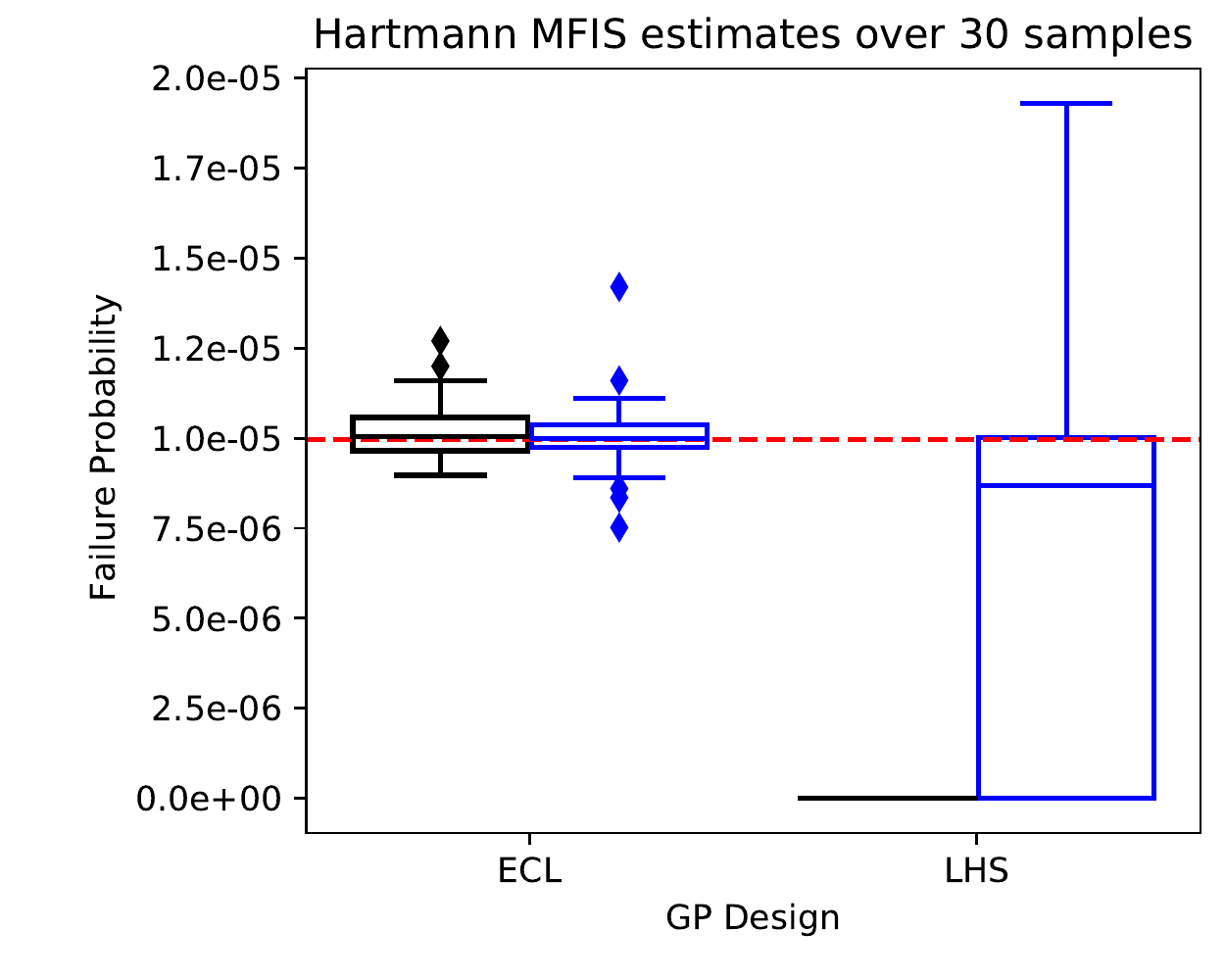}
		\caption{MFIS failure probability estimates $\hat{\alpha}$ for the Ishigami
			(left) and Hartmann-6 (right)  based on GPs trained from ECL adaptive and
			LHS designs over 30 MC repetitions. The color of each box shows the
			definition of $\hat{y}(\x)$ (black: GP mean; blue: GP 95\% upper confidence
			bound) used to classify failure locations for the bias distribution
			$\F_\star$. The true failure probabilities $\alpha$ are shown with the
			red-dashed lines.}
		\label{fig:benchmark_mfis}
	\end{figure*}

	Figure \ref{fig:benchmark_mfis} shows the failure probability estimates
	$\hat{\alpha}$ for 30 repetitions in the Ishigami (left) and Hartmann-6 (right) experiments
	following this scheme. In these problems, the mean surfaces of the LHS-designed GPs rarely exceed the failure threshold, classifying a few (if any) samples as failures, leading to an MFIS estimate of zero. Using the UCB for failure classification often produced enough failures to fit a bias distribution. There is considerable variability in the MFIS estimates based on UCB, but for Ishigami $\hat{\alpha}^\text{Ish}$ is within an order of magnitude of the true $\alpha^\text{Ish}$ (red horizontal dashed line).
	
	Just as in Figure \ref{fig:ad_vs_sf_bakeoff}, using an adaptive design for the
	GP surrogate $\mathcal{S}$ greatly improves the downstream accuracy of MFIS
	estimates. Those estimates are centered around the true values (red dashed
	lines), exemplifying unbiasedness. For the ECL results in Figure \ref{fig:benchmark_mfis},
	using the UCB does not help much, in
	large part due to the much lower uncertainty around the contour thanks to the
	contour targeting in adaptive design. Yet in cases where significant
	uncertainty remains after the adaptive design (LHS results), using UCB may result in a conservative bias distribution over undiscovered sections of the
	failure region(s).
	
	\subsection{Spacesuit impact simulation}
	\label{ss:spacesuit}
\begin{figure*}[ht!]
		\centering
		\includegraphics[trim=0 5 0 10, clip,width=.5\textwidth]{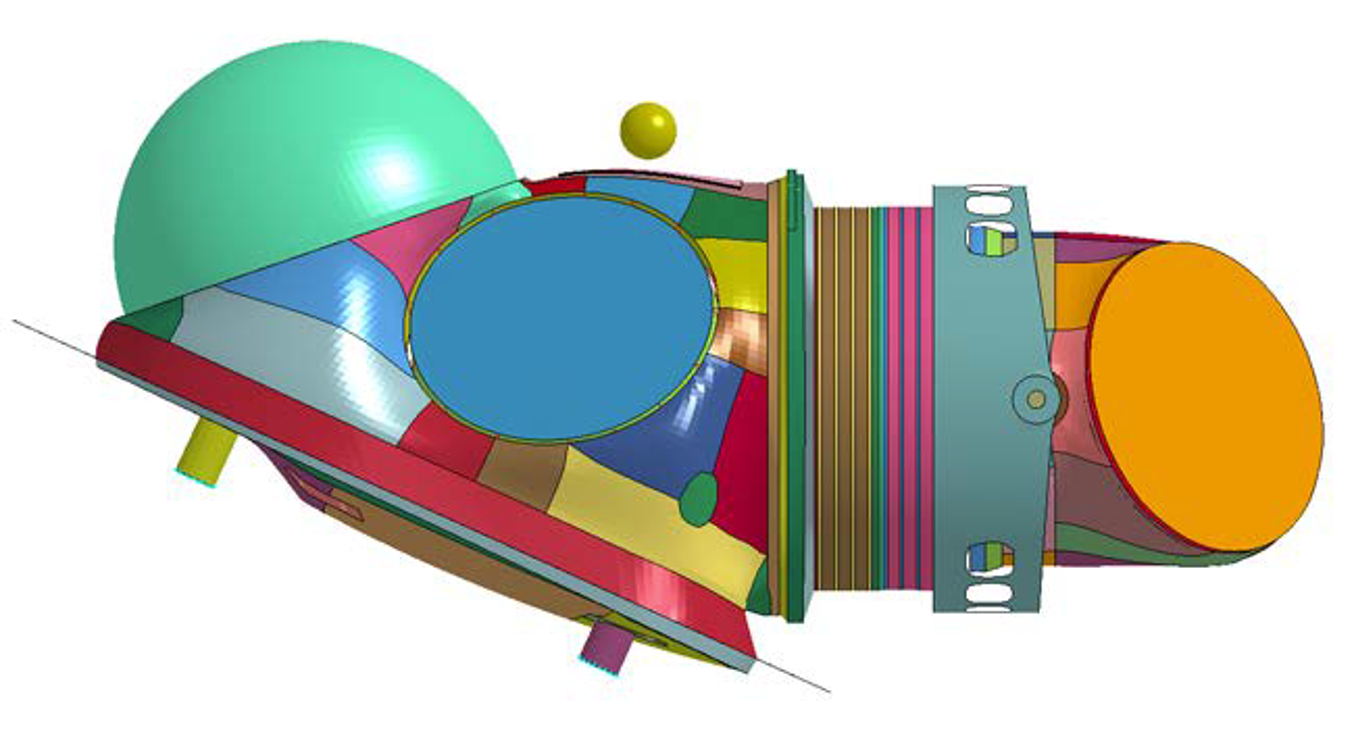}
		\caption{MAT162 assembly model for hard upper torso of Z-2 spacesuit and a rock projectile (yellow sphere).}
		\label{fig:hut}
	\end{figure*}

	NASA's Artemis program is
	centered around a renewed effort of manned space exploration. Within this
	program, NASA is working to develop the next-generation spacesuit, the exploration extravehicular mobility unit (xEMU), to
	provide astronauts with a suit that is robust, lightweight, and mobile. Under
	potential impact loads (due to projectiles, falls, etc.) the suit's
	probability of no impact failure (PnIF) is of top concern for
	certification. With the computational model for the xEMU currently under
	development, we implemented our methodology on a computer model of the previous generation spacesuit, {\em Z-2}, to estimate PnIF under various impact loading
	conditions. The model simulates an impact to the hard upper torso region of the spacesuit (shown in Figure \ref{fig:hut}) using LS-Dyna finite element method software \citep{lsdyna} and the MAT162 material model \citep{gama2009progressive,	haque2017progressive} to capture the resulting progressive damage to the material.
	Based on calibration and sensitivity analysis detailed in \cite{Warner2021}, there are four
	variables in our experiment: three parameters controlling the softening behavior of the material due to damage, henceforth referred to as damage coefficients 1-3, and the impact velocity. The input distribution $\F$
	is $\x \sim \mathcal{N}(\boldsymbol\mu, \boldsymbol\Sigma)$ where
	\begin{equation} \label{eq:input_dist}
		\boldsymbol\mu=
		\begin{bmatrix} 0.41597 \\ 1.54189 \\ 0.01031 \\ 1 \end{bmatrix}
		\; \text{and} \;
		\boldsymbol\Sigma=\begin{bmatrix}
			0.00275 & -0.00494 & -0.00373 & 0 \\
			-0.00494 & 0.01856 & 0.0032 & 0 \\
			-0.00373 & 0.0032 & 0.01834 & 0 \\
			0 & 0 & 0 & 0.00016
		\end{bmatrix}
		. 
	\end{equation}
	The response of interest, $y$, is the maximum contact force, with a threshold of
	2800 lbf. Thus to estimate $\alpha=1-\mathrm{PnIF}$, we define
	$g(y(\x))=y(\x)-2800$. The computer model requires 18 hours on 10 cores to
	produce one sample, severely limiting our ability to entertain a large
	campaign. We budgeted $N=200$ runs for the adaptive design (including an
	initial LHS of 40), with adaptive samples generated in batches of ten so the required simulations could be performed in parallel. For the
	GP surrogate $\mathcal{S}$, we use the Mat\'ern 3/2 \citep{Stein2012} kernel
	with separable lengthscale and a jitter of $10^{-6}$. The data is prescaled to
	$[0,1]^4$ using bounds derived from five standard deviations away from the
	mean in each dimension.

	\begin{figure*}[ht!]
		\centering
		\includegraphics[trim=0 0 0 0 , clip,width=\textwidth]{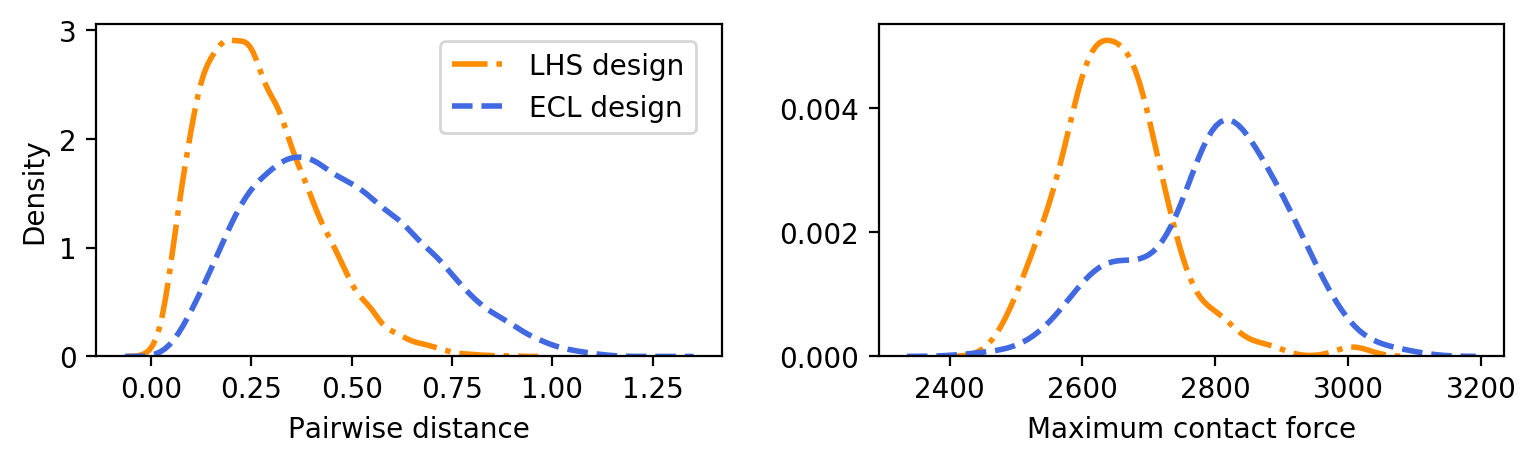}
		\includegraphics[trim=0 0 0 0 , clip,width=\textwidth]{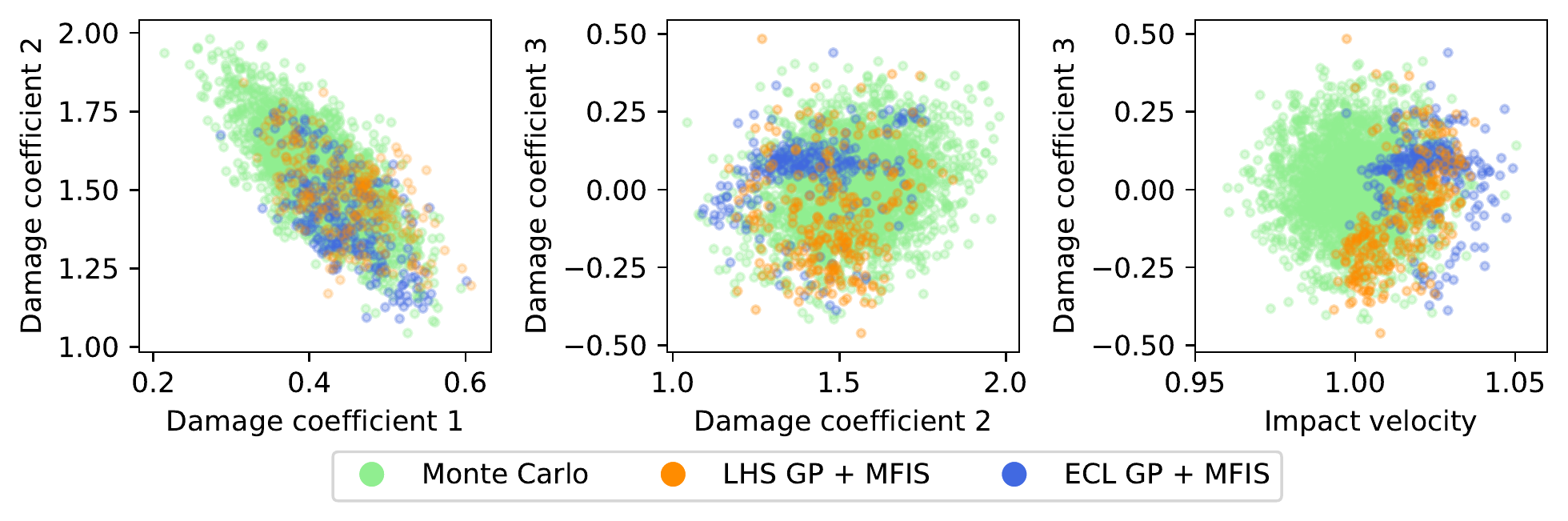}
		\caption{GP designs $(\X_N,\Y_N)$ and MFIS samples $\X_{M^\star}$ for the Z-2
			spacesuit experiment. {\em Top left:} pairwise distances for the inputs.
			{\em Top right:}  observed outputs from GP designs. {\em Bottom:} pairwise plots of samples from input ($\F$, green) and bias ($\F_\star$, orange and blue) distributions used for estimating failure probability.}
		\label{fig:z2_densities}
	\end{figure*}
	
	As a benchmarking comparator another GP was fit to a 200-sized LHS that was
	warped to $\F$ in Eq.~(\ref{eq:input_dist}) via an inverse CDF (dropping the
	covariances). Figure \ref{fig:z2_densities} shows pairwise observed distances
	in both designs' set of inputs. The ECL design contains more long distances
	between points -- the inputs are more spread out in the input space. The
	density in the top right of Figure \ref{fig:z2_densities} shows the focus of
	ECL's samples around $T=2800$.  We take these distinctions as circumstantial
	evidence that our sequential design method is working.
	
	We used $M=10^5$ samples and the $\mathcal{S}$'s UCB to train the bias
	distribution $\F_\star$. Due to the covariances \eqref{eq:input_dist} between
	inputs in $\F$, a Gaussian mixture model with a full covariance structure was
	used for $\F_\star$. We generated $M^\star=250$ samples from both bias
	distributions in order to calculate MFIS estimates. The plots in the bottom of
	Figure \ref{fig:z2_densities} show projections of these samples for select
	pairs of input variables. While the densities of points look similar for
	damage coefficient 2 compared to damage coefficient 1, the ECL MFIS samples
	are biased higher for damage coefficient 3 and impact velocity than compared
	to LHS MFIS samples. This confirms intuition that higher impact velocity
	produces more damage, and thus produces failures. As a reference, a set of
	$M=2500$ pure-MC samples is also shown (generated from $\F$). Both sets of
	samples from bias distributions show a clear tendency towards sections of the
	input space targeting predicted failure region(s).
	
	\begin{figure*}[ht!]
		\centering
		\includegraphics[trim=0 10 0 10 , clip,width=.65\textwidth]{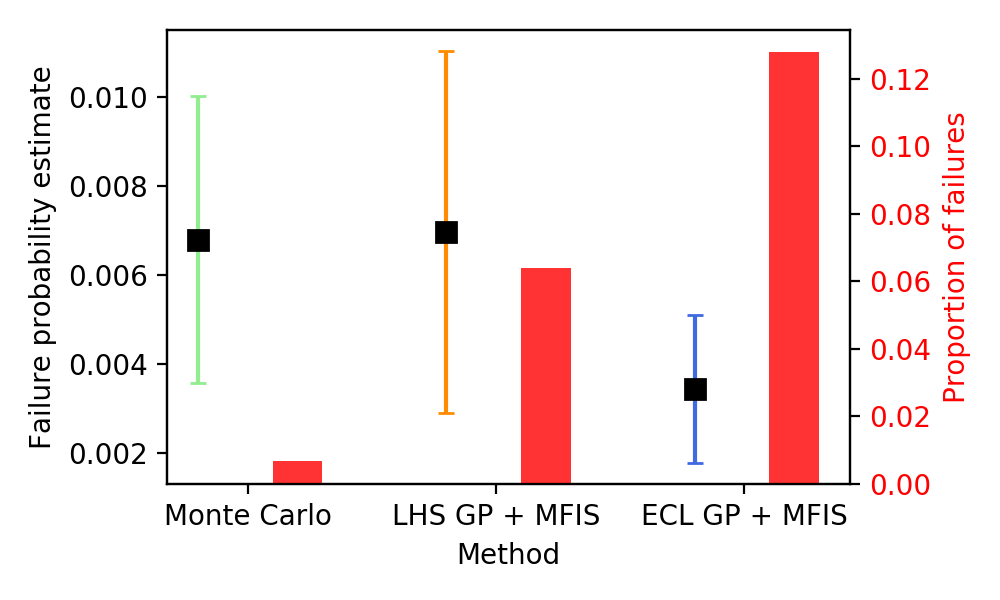}
		\caption{Failure probability estimates $\hat{\alpha}$ with 95\% confidence
			intervals. Red bars show the proportion of samples with failure outputs that
			were used in those estimates.}
		\label{fig:z2_results}
	\end{figure*}
	
	MFIS estimates based on the samples in Figure \ref{fig:z2_densities} are shown
	in Figure \ref{fig:z2_results}. Notice that the 95\% confidence intervals of
	all three estimates overlap, with the ECL GP $+$ MFIS giving the smallest
	point estimate. The ECL GP $+$ MFIS estimate also contains the narrowest
	confidence interval, due to a higher proportion of failures observed in the
	MFIS samples. The proportion of failures observed in both MFIS estimates may
	seem low and are no doubt affected by using the GP UCB for classification during
	$\F_\star$ fitting. A higher failure rate for ECL GP $+$ MFIS samples suggests
	that the bias distribution produced by the UCB of the ECL GP includes the
	unknown true failure region as a larger proportion of the bias distribution's
	density.
	
	\section{Discussion}
	\label{sec:discussion}
	Estimating reliability in engineering applications often entails expensive simulation, limiting the ability to gather large
	quantities of evaluations. Multifidelity importance sampling is one
	strategy to reduce the amount of data needed for an unbiased reliability
	estimate, leveraging a surrogate model (like a GP) and bias distribution to
	reduce estimator variance. If the GP's predictive surface is inaccurate in the
	vicinity of the true failure contour, misclassified samples may be used to fit
	a bias distribution, threatening the overall accuracy of the reliability
	estimate. By marrying a GP built on an adaptive design for contour location
	with MFIS, we can guard against such pitfalls.
	
	We proposed a thrifty adaptive design with a novel criteria: the Entropy-based
	Contour Locator (ECL). Our ECL acquisition function applies the basic
	definition of entropy on events derived from predicted pass/fail from the
	surrogate. Everything for ECL has a closed form.  We
	solve for a new acquisition, maximizing ECL, in such a way that multiple
	failure regions based on high quantiles may be identified. 
	Results showed that our ECL adaptive designs exhibited a degree of
	stochasticity in optimization that promoted exploration of local maxima. When
	problems contain multiple failure regions, especially ones with small relative
	volumes, this exploration is key.  Our examples include illustrative
	benchmarks with two (2d and 6d) and six (3d) failure regions. Against a litany
	of contemporary adaptive design benchmarks, ECL performs among the best at
	correctly identifying true failures as well as estimating the contour's
	volume. All this is accomplished at a one to two orders of magnitude
	computational savings. We also implemented ECL design plus MFIS on spacesuit
	damage simulator, demonstrating how combining these two methods provides a
	point estimate with less uncertainty than plain MC or MFIS techniques.
	
	We demonstrated how the adaptive design process with ECL may be extended to
	batch selection, providing computational savings when parallel-processing is
	available. By selecting multiple points before observing more responses, the
	batch option is more sensitive to the quality of surrogate, and in particular
	its contour estimate(s). Batch size also plays a role in the quality of the
	information gained from the sample points. In high-dimensional problems, batch selection is crucial for the method to be tractable. While we fixed the batch size for our experiments, the method could likely be improved by setting the batch size as a function of available computing time.
	
	Our methodological contribution focused on targeting failure regions with ECL,
	not explicitly on its interface with MFIS. Rather, we treated MFIS as one, of
	possibly several, downstream reliability tasks after ECL acquisition.  This
	compartmental nature is an asset, as it means our contribution is portable.
	However, it also means potential for improved efficiency in the reliability
	context was left on the table. We believe there is opportunity for further
	blending of adaptive design and MFIS: perhaps by generating acquisitions and
	refining MFIS bias distributions $\F_\star$ at the same time.

	\bibliographystyle{jasa}
	\bibliography{entropy_mfis_articles}
	
	\appendix
	\section{Value of integration knot density}
	\label{app:grid_density}
	
	When applying any of the adaptive designs in Section \ref{ss:syn_results},
	there are certain choices, or ``knobs", that play a crucial role in optimizing
	the acquisition function. Though both CLoVER and ECL use entropy in their
	acquisition functions, performing adaptive design with CLoVER includes two
	knobs, whereas ECL only has one. Here we explore the effect of different knob
	settings for experiments detailed in Section \ref{sec:toy_results}.
	
\begin{figure*}[ht!]
		\centering
		\includegraphics[trim=0 0 33 28, clip,width=0.49\textwidth]{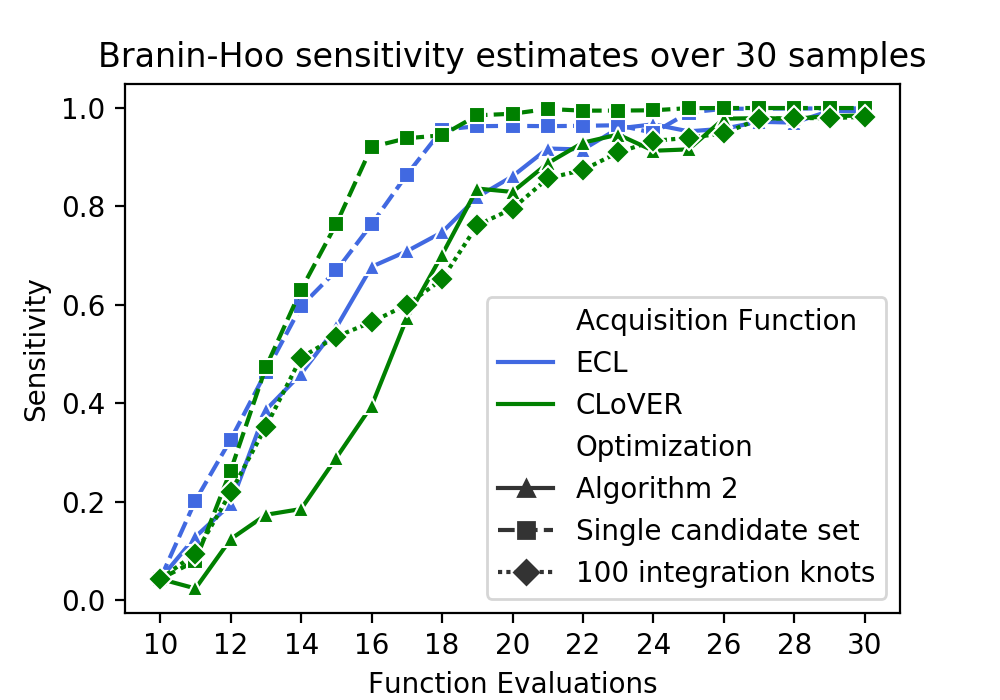} 
		\includegraphics[trim=0 0 33 28, clip,width=0.49\textwidth]{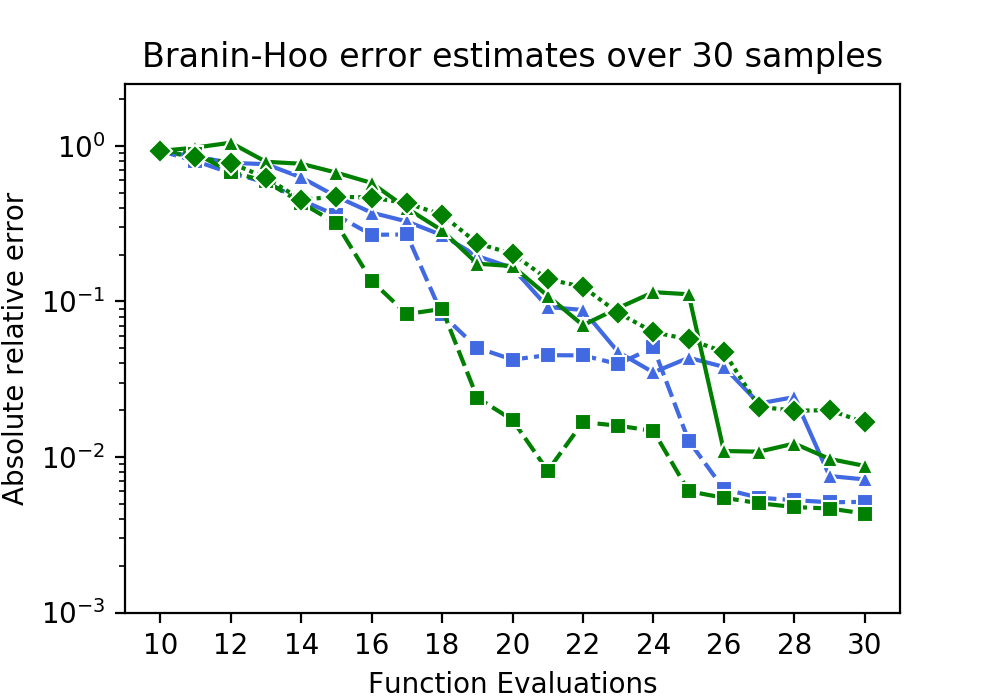}
		\caption{Mean sensitivity (left) and relative error of the volume estimate
			(right) for the Branin-Hoo experiment over 30 MC repetitions. The ECL (blue)
			and CLoVER (green) acquisition functions are used with different
			optimization strategies.}
		\label{fig:branin_grid_density}
	\end{figure*}

	To begin, we focus on the strategy used to optimize the acquisition function.
	ECL uses Algorithm \ref{alg:entropy_opt}, which contains a small set of $N_c =
	10d$ candidate points and a continuous optimization initialized at the best
	candidate. For CLoVER, \cite{marques2018contour} propose candidates only,
	forgoing a continuous optimizer. In their code, they use a single candidate
	set from which to select all new design locations. For our experiments, we fix
	that set to be an LHS of 1000 points. The results shown in Section
	\ref{ss:syn_results} use these strategies. In Figure
	\ref{fig:branin_grid_density}, we supplement that experiment with swapped
	optimization strategies for ECL and CLoVER. ECL's original strategy is denoted
	with solid lines and triangles and CLoVER's single candidate set strategy is
	shown with the dashed lines and squares. CLoVER and ECL follow a similar
	trajectory for sensitivity and relative error when a single candidate set is
	used. With Algorithm \ref{alg:entropy_opt}, CLoVER's sensitivity lags behind
	ECL during the first half of the experiment.
	
\begin{figure*}[ht!]
		\centering
		\includegraphics[trim=0 0 30 28, clip,width=0.49\textwidth]{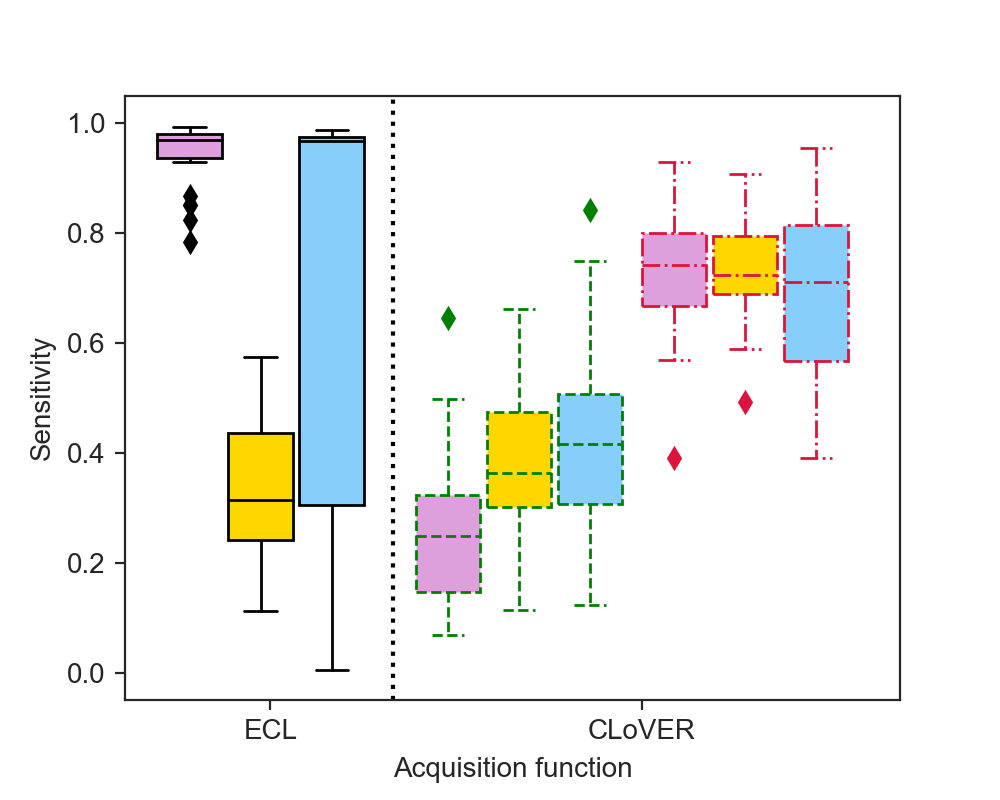} 
		\includegraphics[trim=0 0 30 28, clip,width=0.49\textwidth]{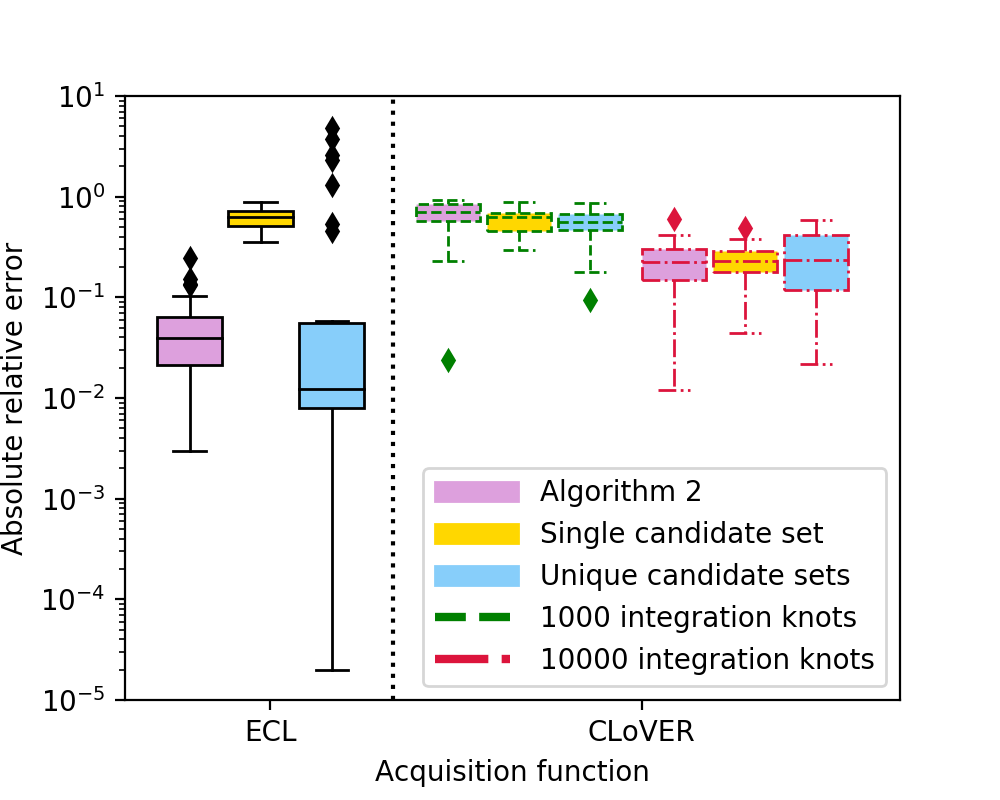}
		\caption{Final sensitivity (left) and relative error (right) for the Ishigami experiment. Different optimization strategies are used on the ECL and/or CLoVER acquisition functions, changing the number of candidate points and integration knots.}
		\label{fig:ishigami_grid_density}
	\end{figure*}

	In the Ishigami experiment, we find a more complex story (Figure \ref{fig:ishigami_grid_density}). Here we provide the sensitivity and absolute relative error distributions for the 30 repetitions at the experiment's conclusion. Since the number of sequential samples (170) is a large proportion of the candidate set's size, we introduce a third optimization strategy that generates unique 1000-sized LHSs at each adaptive step (blue boxes). For ECL, using Algorithm \ref{alg:entropy_opt} for optimization (purple box) clearly beats both a single candidate set (yellow box) or unique candidate sets for each iteration. 
	
	Unlike with ECL, the CLoVER results in Figure \ref{fig:ishigami_grid_density} do not show a drastic change in sensitivity or relative error across optimization strategies. With the Ishigami experiment being harder
	(larger quantile, more failure regions) than Branin-Hoo, the original set of 1000
	integration knots may be too small. Increasing to 10000 knots shows a
	significant improvement for all optimization strategies. Looking back at the
	Branin-Hoo experiment in Figure \ref{fig:branin_grid_density}, we added a
	CLoVER variation using a single candidate set and only 100 integration points
	(dotted line with triangles). We find the sensitivity is below CLoVER with
	1000 knots for a majority of the experiment and the relative
	error is significantly higher.
	
	\begin{table}[h!]
		\centering
		\begin{tabular}{|l|r|r|r|} 
			\hline
			\multirow{2}{*}{\bfseries Optimization Strategy} & 
			\multirow{2}{*}{\bfseries ECL} & 
			\multicolumn{2}{c|}{\bfseries CLoVER}\\ \cline{3-4}
			&& $10^3$  knots & $10^4$  knots \\ \hline
			Algorithm \ref{alg:entropy_opt} & 0.008  & 15.2 & 17.1  \\ \hline
			Single candidate set & 0.09 & 17.8 & 32.4 \\ \hline
			Unique candidate sets & 0.14 & 18.2 & 28.7 \\
			\hline
		\end{tabular}
		\caption{Average computation times (minutes) to select points across 30 MC
			repetitions of the Ishigami experiment.}
		\label{tab:ish_timings}
	\end{table}
	
	Increasing the number of optimization candidate points or integration knots
	comes at a price -- namely computation time. Table \ref{tab:ish_timings}
	summarizes the drastic difference in computing times between ECL and CLoVER for
	the Ishigami experiment. For challenging problems like Ishigami, an even
	greater number of integration knots seems necessary to produce competitively
	accurate results. When accounting for the additional computational burden, implementing CLoVER with a large number of knots in higher dimensional problems (i.e. Hartmann-6) could be impractical based on current software.

\end{document}